\pdfoutput=1
\PassOptionsToPackage{hyphens}{url}
\documentclass[conference]{IEEEtran-mod} 

\usepackage{enumerate}
\usepackage{graphicx}
\usepackage{epsfig}
\usepackage{color}
\usepackage{url}
\usepackage{etoolbox}
\usepackage{booktabs}
\usepackage{amsmath}
\usepackage{theorem}
\usepackage{amssymb}
\usepackage{mathtools}
\usepackage{multicol}
\usepackage{multirow}
\usepackage{hhline}
\usepackage{setspace}
\usepackage{balance}
\usepackage{caption}
\usepackage{xcolor,colortbl}
\usepackage{array}
\usepackage{xspace}
\usepackage[normalem]{ulem}
\usepackage[colorlinks]{hyperref}
\hypersetup{
	linktoc=page,     
	linkcolor=[rgb]{0.4,0.3,1},  
	urlcolor=black,
	citecolor=[rgb]{0,0.6,0.4}
}

\usepackage{amssymb}
\usepackage{pifont}

\definecolor{Gray}{gray}{0.9}
\definecolor{persiangreen}{rgb}{0.0, 0.65, 0.58}
\definecolor{viridian}{rgb}{0.25, 0.51, 0.43}
\definecolor{ruddy}{rgb}{1.0, 0.0, 0.16}

\newcommand{\cmark}{\color{persiangreen}\ding{51}}%
\newcommand{\xmark}{\color{ruddy}\ding{55}}%

\newcommand{\system}{LPAuditor\xspace}

\newcolumntype{x}[1]{>{\centering\arraybackslash\hspace{0pt}}p{#1}}

\newcommand{\tabitem}{~~\llap{\textbullet}~~}

\newenvironment{squishlist}
{\begin{list}{$\bullet$}
		{ \setlength{\itemsep}{1pt}
			\setlength{\parsep}{1pt}
			\setlength{\topsep}{1pt}
			\setlength{\partopsep}{1pt}
			\setlength{\leftmargin}{1.5em}
			\setlength{\labelwidth}{1em}
			\setlength{\labelsep}{0.5em} } }
	{\end{list}}

\begin{document}

\title{Please Forget Where I Was Last Summer:\\ The Privacy Risks of Public Location (Meta)Data}

\author{\makebox[.99\linewidth]{
		Kostas Drakonakis,\IEEEauthorrefmark{1}
		Panagiotis Ilia,\IEEEauthorrefmark{1}
		Sotiris Ioannidis,\IEEEauthorrefmark{1}
		Jason Polakis\IEEEauthorrefmark{2}}\\

	\and \makebox[.01\linewidth]{} 
	\and
	
	\IEEEauthorblockA{\IEEEauthorrefmark{1}
		FORTH, Greece\\ 
		\{kostasdrk, pilia, sotiris\}@ics.forth.gr}
	
	\and
	\IEEEauthorblockA{\IEEEauthorrefmark{2}
		University of Illinois at Chicago, USA\\
		polakis@uic.edu}
}

\IEEEspecialpapernotice{\vspace{0.5em} \large{\color{blue}This is an extended version of our paper that will appear at NDSS 2019}}
	
\maketitle

\begin{abstract}
The exposure of location data constitutes a significant privacy risk to users as it
can lead to de-anonymization, the inference of sensitive information, and
even physical threats. In this paper we present \system, a tool that conducts a 
comprehensive evaluation of
the privacy loss caused by publicly available location metadata.
First, we demonstrate how our system can pinpoint
users' key locations at an unprecedented granularity by identifying their
actual postal addresses. Our experimental evaluation on Twitter data highlights the
effectiveness of our techniques which outperform prior approaches by
18.9\%-91.6\% for homes and 8.7\%-21.8\% for workplaces. Next we present a novel exploration 
of automated private information inference that uncovers ``sensitive'' locations that
users have visited (pertaining to health, religion, and sex/nightlife). We find
that location metadata can provide additional context to tweets and thus lead to
the exposure of private information that might not match the users' intentions.

We further explore the mismatch between user actions
and information exposure and find that older versions of the official
Twitter apps
follow a privacy-invasive policy
of including precise GPS coordinates in the metadata of tweets that
users have geotagged at a coarse-grained level (e.g., city).
The implications of this exposure are further exacerbated by our finding that
users are considerably \emph{privacy-cautious} in regards to exposing
precise location data. 
When users can explicitly select what
location data is published, there is a 94.6\% reduction in tweets with GPS
coordinates. As part of current efforts to give users more control over their data,
\system can be adopted by major services and offered as 
an auditing tool that informs users about sensitive information
they (indirectly) expose through location metadata.
\end{abstract}

\section{Introduction}
\label{sec:introduction}

The capability of modern smartphones to provide fine-grained 
location information in real time has enabled 
the deployment of a wide range of novel functionality by online services.
In Twitter users can
incorporate location information in their tweets to provide more context and enrich 
their communications~\cite{Patil:2012}, or even enhance situational awareness during 
critical events~\cite{Vieweg2010}.
Nonetheless, the presence of location metadata in a by-default-public data stream
like Twitter constitutes a significant privacy risk. Apart from potentially enabling physical threats
like stalking~\cite{gross2005information,polakis:ccs15} 
and ``cybercasing''~\cite{Friedland:2010}, location information could lead to the inference 
of very sensitive data~\cite{Minami2010,blumberg2009}, 
and even get combined with other information collected from online services~\cite{polakis:badgers15}.
Previous work has demonstrated how to identify users' key locations (i.e., home and work)
at a postcode~\cite{efstathiades_2015} or very coarse-grained ($\sim$10,000m\textsuperscript{2}) 
level~\cite{hu_2015,cheng2011exploring}. However, this coarse granularity 
fails to highlight the true extent of the privacy risks introduced by the public availability
of geographical information in users' tweets. Furthermore, 
these studies have not explored what sensitive information can be inferred from users geotagging
tweets at other locations.

In this paper we develop \system, a system that examines the privacy risks
users face due to publicly accessible location information, and conduct a large scale study
leveraging Twitter data and public APIs. Initially 
we present techniques for identifying a user's home and work
at a postal address granularity; our heuristics are built around intuitive social and behavioral norms.
We first conduct a two-level clustering process for creating clusters of
tweets and mapping them to postal addresses, which is robust to errors in the
GPS readings of smartphones~\cite{zandbergen2009accuracy} as well as minor
spatial displacement due to user mobility (e.g., the user tweeting while
arriving or departing from home). Subsequently we analyze the spatio-temporal
characteristics of a user's tweets and break down the user's tweeting behavior
in time windows of varying granularity; we identify the most long-lived
clusters in regards to weekend tweeting activity, and then select the one with
the largest number of one-hour time windows that contain at least one tweet.
For identifying users' workplaces, we introduce the first adaptive technique
that handles users with varying or split shift schedules at work. Specifically,
we identify the longest clusters in terms of active weeks. Then,
we extract the \emph{dominant time frame} of each cluster by superimposing
their daily time frames, even ones spanning two different days (i.e. night shifts),
and excluding any hours that appear to be outliers, as well
as any clusters that have a repeated activity of more than 10 hours and, finally, 
select the cluster with the largest number of active weeks.

Through an arduous manual process we create a ground truth dataset for 2,047 users,
which enables us to experimentally evaluate our auditing tool. Our
system is able to identify the home and workplace for 92.5\% and 55.6\% of the users respectively.
When compared to state-of-the-art results, we find that our techniques outperform previous
approaches by 18.9\%-91.6\% for homes and 8.7\%-21.8\% for workplaces.

Apart from the increased effectiveness of our techniques, our work demonstrates that
by leveraging widely available geolocation databases attackers can pinpoint users' key locations
at a granularity that is orders of magnitude more precise than previously demonstrated.
Without doubt, this level of accuracy renders the identification of users a trivial task.
The privacy implications of our findings are even more alarming when considering
the prominent role that platforms like Twitter play in 
protests and other forms of social activism~\cite{huang2011facebook}.
Indeed, a substantial number of users choose to not reveal their 
actual identity on Twitter, and prior work has found a correlation between the choice of anonymity and 
the sensitivity of topics in tweets~\cite{Peddinti2014}\footnote{This study also reported that 
5.9\% of Twitter accounts are anonymous and another 20\% do not disclose their full name.} 
and other online posts~\cite{Peddinti2014_swagger}.

\system offers a comprehensive analysis of the privacy loss
caused by location metadata by also exploring whether the remaining locations
can be used to infer personal information that is typically 
considered sensitive. While the inference of sensitive information has been one of 
the main motivations behind prior research on location-privacy~\cite{primault}, such automated 
attacks have not been demonstrated in practice.
Our system examines tweets that place the user at (or in close
proximity of) locations that are associated with such information. Currently we search for 
locations pertaining to three sensitive topics: religion, medical issues, and sex/nightlife.
We find that 71\% of users have tweeted from sensitive locations,
27.5\% of which can be placed there with high confidence
based on the content of their tweets. 
Privacy loss is amplified by the location metadata as it 
leaks additional contextual details to the tweet's content;
e.g., the user may simply mention
being at a doctor without giving more details, while the location metadata
places the user at an abortion clinic.
We also explore a spatiotemporal-based
approach and find that 29.5\% of the users can be placed at a sensitive location
regardless of the content of their tweets.
As such, we envision \system being
offered as an auditing tool by social networks and location-based
services, providing users with an overview
of the sensitive information that can be inferred based on their
publicly accessible location data. This can support
recently stated initiatives of giving users more control over their data~\cite{reuters_facebook}.

Finally, our study reveals that older versions of the Twitter app implement a privacy-invasive 
policy. 
Specifically, tweets that are geotagged by users at a coarse
level (e.g., city) include the user's exact coordinates in the tweets' metadata. This privacy 
violation is invisible to users, as the GPS coordinates are only contained in the metadata returned by 
the API and not visible through the Twitter website or app. To make matters worse, 
this historical metadata \emph{currently remains publicly accessible}
through the API. We quantify the impact of Twitter's invasive policy,
and find that it results in an almost 15-fold increase in the number of users
whose key locations are successfully identified by our system. 
In an effort to remediate this significant privacy threat we have disclosed our findings to Twitter.
In summary, our main research contributions are:

\begin{squishlist}

\item We conduct a comprehensive, IRB-approved, large-scale exploration of the privacy
risks that users face when location data is, either explicitly or
inadvertently, shared in a public data stream like Twitter's API.

\item We develop \system, a system that leverages location metadata for
identifying key locations with high precision, outperforming
state-of-the-art approaches. Apart from achieving superior granularity,
we also introduce a clustering approach that renders our system robust to 
errors in GPS readings or spatial displacement due to user mobility.

\item We present the first, to our knowledge, study on the feasibility of \emph{automated}
location-based inference attacks. Our system leverages
novel content-based and spatiotemporal techniques for inferring sensitive user information, thus,
validating the motivation of prior location-privacy research.

\item We measure the impact of Twitter's invasive policy for collecting and sharing precise location data
and quantify the lingering implications. Our study on user geotagging behavior reveals that users are
restrained when publishing their location and avoid including exact coordinates when given control by 
the underlying system, yet remain exposed due to the availability of this historical data.

\end{squishlist}

\section{Motivation and Threat Model}
\label{sec:motivation}

The sensitive nature of mobility data is well known to the research
community, which has proposed various techniques so far for limiting the granularity
of the location data that services can obtain (e.g.,~\cite{Gruteser:2003}). In
practice, however, such defenses have not seen wide deployment and a large
number of mobile apps collect precise locations~\cite{Ren:2016}. While
prior work has proposed approaches for identifying key locations (home and
work), the reported granularity is not sufficient for demonstrating the true
extent of the threat (e.g.,~\cite{efstathiades_2015,hu_2015,cheng2011exploring,
cho_2011}). More importantly, the risk of sensitive information being
inferred from other location data points remains unexplored.

Despite the privacy risk this data poses to users, services do not stringently
prohibit access to it and may expose it to third parties~\cite{Krishnamurthy:2010}
or render it publicly accessible. 
To demonstrate the
extent and accuracy of sensitive information inference that an adversary can achieve,
we develop and evaluate \system exclusively using public and
free data streams and APIs. Furthermore, we design our system to be application-independent
and applicable to other location datasets.
We show that location metadata
enables the inference of sensitive information that could be misused for a wide
range of scenarios (e.g., from a repressive regime de-anonymizing an activist's
account to an insurance company inferring a customer's health issues,
or a potential employer conducting a background check). While we 
build a tool that can be adopted by online services for better protecting users' privacy,
the techniques employed by our system could be applied by a wide range of
adversaries or invasive third parties. By demonstrating the severity 
and practicality of such attacks, we aim to initiate a public discussion
and incentivize the adoption of privacy-preserving mechanisms.

\section{System Overview}
\label{sec:methodology}

In this section we provide an overview of our system.
First we describe how \system clusters location data
and identifies key locations. Next
we provide details on our methodology for identifying sensitive locations that users may have
visited.
Finally, we provide some implementation details.

\subsection{Data Labeling and Clustering}

\textbf{Labeling tweets.}
The first step is to label each geotagged tweet with 
the corresponding postal address. To highlight the extent of the risk that users face, we 
opt for publicly available API services that could be trivially employed by attackers
for mapping each tweet's GPS coordinates to an address.
To that end, we use the reverse geocoding API by ArcGIS~\cite{arcgis_api} for the majority
of our labels, and the more accurate but rate-limited Google Maps Geocoding API~\cite{google_api} 
for the subset of labels that are more critical to our accuracy.
While this allows us to improve our system's performance, in practice,
if \system is adopted by a major service like Twitter, Google Maps could
be used for the entirety of the calls.

Since our dataset is large in size, we developed a form of \emph{caching} that
allows avoiding unnecessary API calls. 
Instead of issuing a call for every pair of coordinates we come across,
we estimate the spatial position of the pair of coordinates and
search for nearby coordinates that have already been labeled.
If the distance to a labeled pair of coordinates is less than two meters,
we assign the same address label to the new pair of coordinates.
Experimentally, we found that this approach reduced the number of API calls 
our system issued by 42.5\%. 

It should be noted, however, that
geocoding APIs do not always return an address. 
We label those tweets with ``unknown address''.
After a manual investigation and verification of a random subset,
we observed that they typically correspond to 
places 
like university
campuses, airports or remote rural areas that do not have 
exact postal addresses. Nonetheless, while we don't have a postal address
in these cases, the granularity of our process is unaffected as we still
have the GPS coordinates.

\textbf{Initial clustering.} \system groups
tweets assigned to the same postal address
into a single cluster (we refer to this as first-level clustering). Then, by taking 
into consideration the coordinates of all the tweets of a cluster, we calculate 
the coordinates for the cluster's mid-point (geometric center).
To verify that the label assigned to a cluster indeed corresponds to 
the cluster's actual address, we use the Google Maps API for retrieving 
the address of the cluster's mid-point coordinates. If the address returned 
from Google's API does not match the already assigned address label for 
the cluster, due to incompatibilities between the two APIs or
borderline cases where our \textit{caching} approach
results in assigning a neighboring address, we opt for the address returned 
from Google's API. However, due to Google's
stricter API rate limits, we only use this methodology for verifying the address of 
the 10 largest clusters of each user, which we have empirically found to be the 
most prominent and significant ones. This follows our threat model 
constraint of demonstrating what attacks can be conducted using free and public APIs.
In practice, attackers with many resources could avoid rate limiting or use
other proprietary geolocating databases.

For tweets with the ``unknown address'' label we
employ the 
DBSCAN algorithm~\cite{dbscan}. 
We empirically set our threshold to 30 meters, but due to its cascading effect
we may cluster together points that have a greater distance 
due to other points laying in between them.
We only use DBSCAN for clustering tweets that have been marked with 
``unknown address'' ($\sim$16\% of clusters);
nearby tweets that have been labeled with 
an actual address are not considered by DBSCAN.

\textbf{Second-level clustering.} We have observed that the initial 
clustering approach can result in multiple neighboring clusters 
for a specific place. The most common case involves one large 
dominant cluster in the area and a few significantly smaller 
clusters next to it, in close proximity.
In general, it is difficult to distinguish which 
tweets belong to each cluster, even by plotting the coordinates 
of these tweets on a map and visually inspecting them.
Through an empirical analysis, where we visually inspected 
clusters and cross referenced the timing of their tweets,
it became apparent that these closely neighboring clusters
typically correspond to a single user location but have been
mapped to a neighboring address. 
Various factors can lead to this,
such as inaccuracies in the user's GPS readings~\cite{zandbergen2009accuracy},
the precision of the geocoding APIs, as well as differences in the
actual tweeting position of the user (tweeting when
leaving a place or arriving, being in the backyard or at a
neighbor etc.).

\begin{figure}[t]
\centering
	\includegraphics[width=0.9\columnwidth,height=3.8cm]{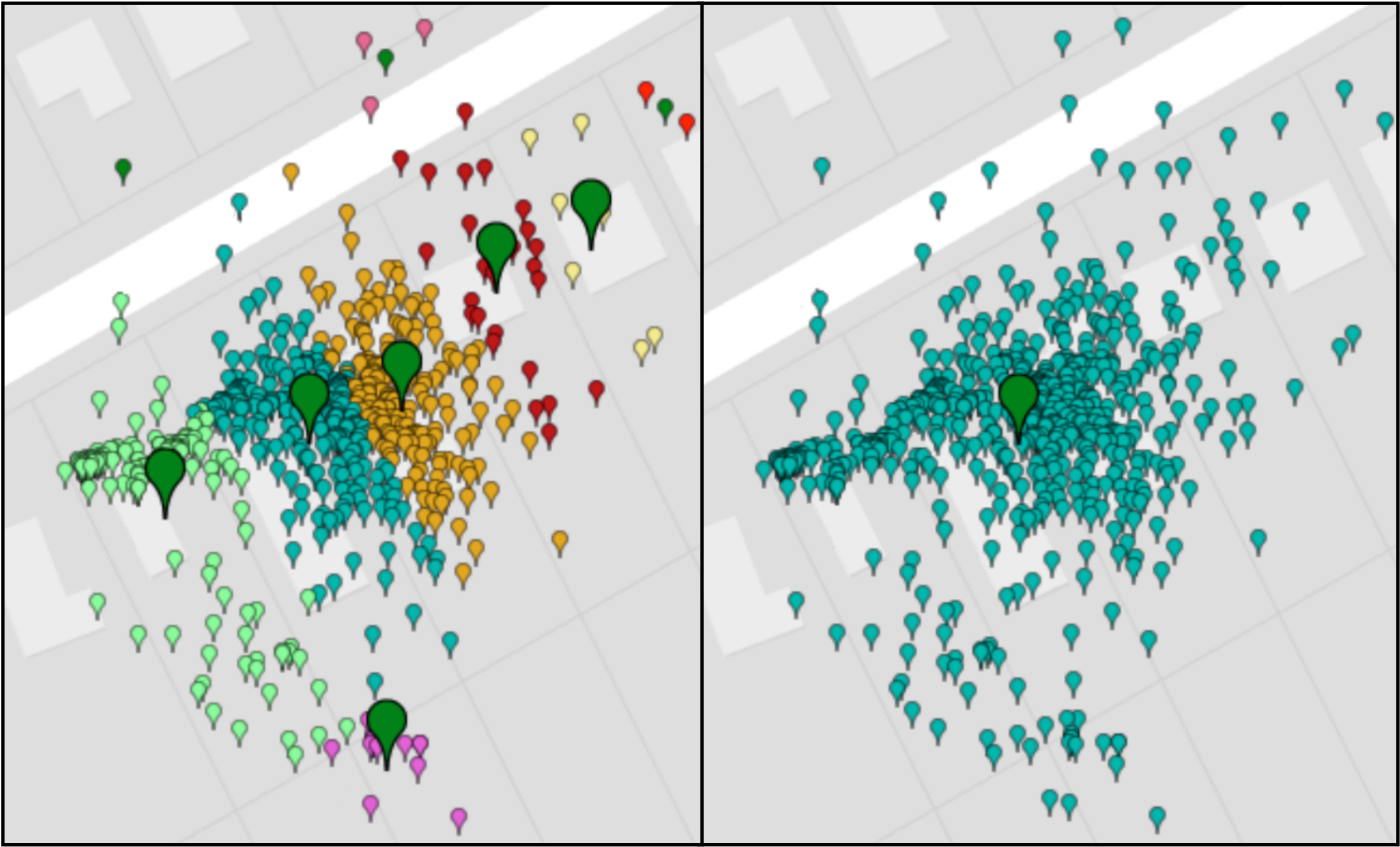}
	\caption{Example of our second-level clustering.}
	\label{fig:collapse}
\end{figure}

As these nearby clusters most likely correspond to the same
place, 
we implement a second-level clustering for grouping 
such neighboring clusters into a larger one.\footnote{For the 
remainder of the paper, referring to \emph{clusters} will imply \emph{second-level clusters}
unless stated otherwise.} 
First we identify which cluster in an area of multiple neighboring
clusters is dominant, i.e., has the most tweets,
and then we employ a modified version of the DBSCAN
algorithm for estimating which clusters should be merged with the
dominant one. For this clustering we consider that the distance
between the mid-point of the larger central cluster and all the smaller
ones should not exceed 50 meters.\footnote{We set this threshold 
based on the value in the FCC mandate for 911 caller location accuracy~\cite{fcc},
as it can account for GPS errors but is not prohibitively large so as to lead to false positives.
We also experimentally verified its suitability.}
To eliminate DBSCAN's cascading effect we check this distance before deciding
whether a cluster should be included in the new one. 
An example
is shown in Figure~\ref{fig:collapse}, where there are several clusters of 
significant size, with smaller clusters around them. All tweets
in a cluster are depicted with a common color, and
the dark green pins denote the center of the various clusters, with the size
of each cluster pin being dependent on the number of tweets mapped to it.

Overall, implementing our second-level of clustering allows us to introduce a
(configurable) radius for effectively mapping these ``runaway'' data points to
the main cluster. Nonetheless, it is important to note that the initial clustering
step (using the geocoding API) is actually necessary; solely applying DBSCAN's
radius-based clustering to the dataset leads to oversized clusters and
eliminates the finer granularity that is achieved by the two-level clustering
approach.

\subsection{Identifying Key User Locations}
\label{sec:key_inference}

Here we describe how \system
selects the clusters that represent two key user locations (home and workplace)
in an automated fashion.
Our system does not take into 
consideration the content and semantics of the tweets posted, but only 
the temporal characteristics and distribution of the tweets in each cluster.
It should be emphasized that our work focuses on location metadata and not the tweet content as this allows us to
quantify the true extent of the privacy risks introduced by location metadata: \emph{even
cautious users that do not explicitly disclose information about their key locations
face this privacy loss.} 
However, \system leverages the content for increasing confidence
in placing users in \emph{other} sensitive locations, as discussed in Section~\ref{sec:sensitive}.
It is imperative to note that our system incorporates heuristics that 
are built upon intuitive assumptions regarding common human behavior and legislative norms
(e.g., 8-hour shifts) in the US (location of our study's users) and many other countries as well (e.g., in the European Union).
While highly effective, these heuristics may require tweaking for countries with 
vastly different social norms or legislature; such cases are out of the scope of this work.

First we obtain the local timezone corresponding to each 
cluster from its mid-point coordinates, and then convert the 
timestamp of each tweet to the local time. This allows us to identify 
tweets that have been posted within specific hours,
and understand the user's daily routine and behavior in depth.
For capturing the temporal characteristics of each cluster and 
understanding the user's activity and tweeting patterns, our system
identifies \emph{active time windows}, i.e., time windows with at least one geotagged 
tweet for a particular cluster. Apart from days or weeks, time windows can be set
to represent weekdays, weekends, or even specific time frames at a granularity 
of hours (e.g., afternoon, late night).

\textbf{Homes} exhibit distinct characteristics compared
to other places users regularly visit, as that is where
people typically return at the end of the day and also spend a 
considerable amount of time. Due to the non-ephemeral relationship 
people have with their home, the temporal characteristics of 
a user's tweeting behavior can sufficiently distinguish this location from
other visited locations. One exception could be users that are considerably 
privacy-cautious and refrain from posting geotagged tweets from their home or
surrounding areas.

Our approach for identifying a user's home cluster is based on 
the following intuitions: (i) as the user spends some time at home every 
day, we expect to \textit{repeatedly} observe some activity from this cluster 
(i.e., multiple active windows in this cluster's timespan), and (ii) the tweets 
of this cluster will not occur solely within a specific time frame, but we 
expect tweets that correspond to almost all hours in the day. In other 
words, while other clusters of a user may follow a specific well-defined
temporal pattern, we expect the home cluster to exhibit a more ``chaotic''
behavior in the long term, having tweets that were posted at different 
times throughout the day, from early in the morning to very late at night.

\begin{figure}[t]
\centering
\includegraphics[width=1\linewidth]{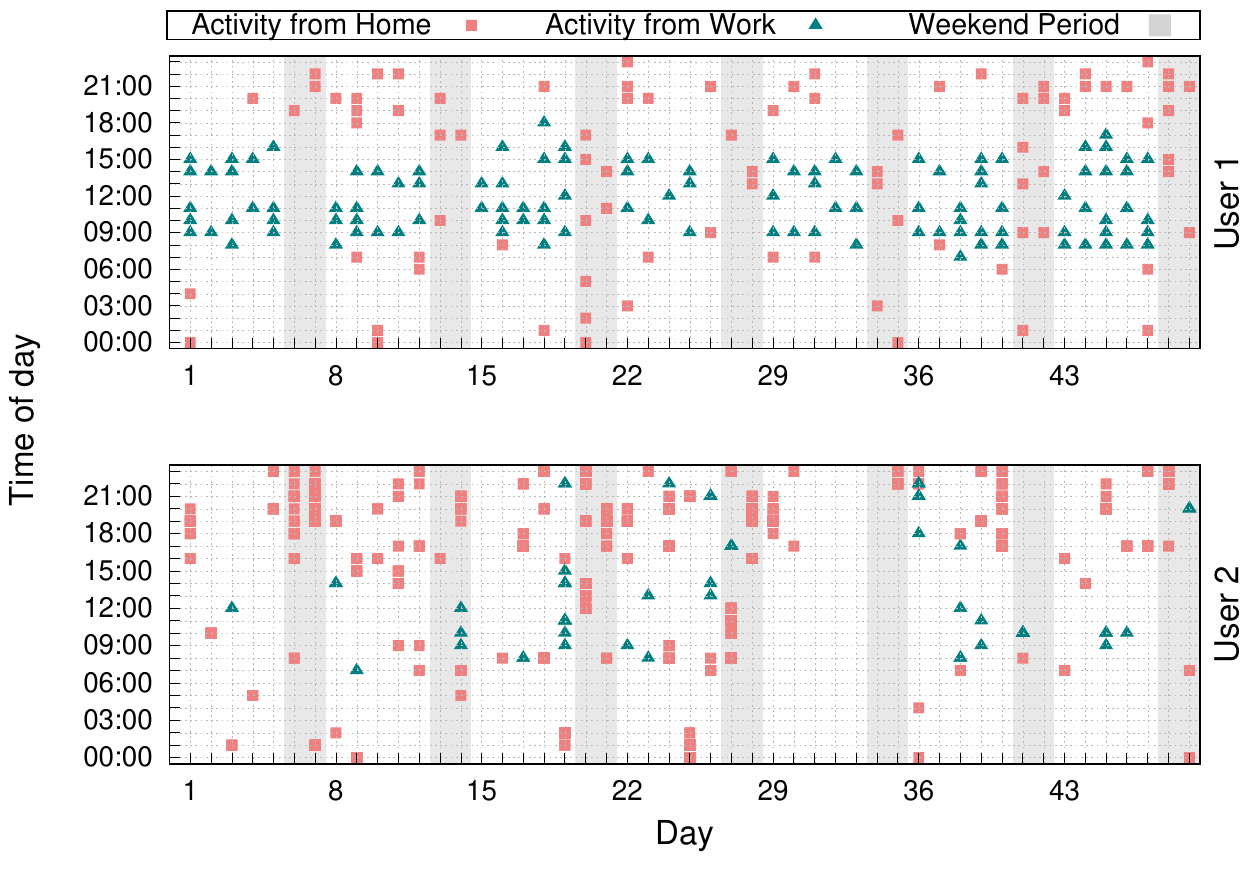}
\caption{An example diagram representing the tweeting activity of two users in our 
dataset from their home and work clusters. User1 exhibits a more
``traditional'' activity pattern, while User2 exhibits erratic
patterns with different work-shifts.}
\label{fig:scaterplot}
\end{figure}

While experimenting with two approaches for specifying the time 
windows (i.e., only weekends vs entire weeks), 
we observed that a week-based time window may introduce 
uncertainty for users that exhibit considerable
activity from multiple clusters. 
As such, we design a robust home-inferring algorithm
by only considering weekends.
We determine which are the user's five most active clusters ``horizontally'',
i.e., those with the highest number of active weekends,
and estimate the time frame and active hours of each of these clusters.
Following our intuition that the home will exhibit more widespread
temporal activity from a macroscopic viewpoint, we choose the cluster 
with the broadest time frame as the user's home.

\textbf{Work.} We expect that, for most users,
tweets posted from work will follow a well-defined time frame
that corresponds to the user's working hours. 
We set the time window to the entire week and identify the five most active 
clusters, i.e., those with the highest number of active weeks (in the 
horizontal dimension). We ignore the home cluster when assembling this set.
For each of the remaining candidate clusters we try to identify the cluster's
most dominant time frame. To that end, we identify all the distinct days
in which the user has posted more than one tweet, and use the
day's earliest and latest tweet for calculating the time frame of that day. After 
estimating the time frame of each active day, we superimpose all
these time frames and consider as the dominant time frame the set of hours
that appears in more than half of the active days of the cluster.  This allows
us to avoid including insignificant 
hours, e.g., for days where the user happened to go to work a
little earlier or later than usual. At the same time this also allows us 
to handle users that have a more lax schedule or may work in shifts. 
We also account for users that work night shifts, which span two consecutive dates;
specifically, we consider instances of active time windows that span 
two days, have a duration of up to eight hours~\cite{labor_shifts} 
and terminate by 07:00,\footnote{In the United States the night shift is
typically 23:00-07:00 while the European Union identifies it as including the
00:00-05:00 period~\cite{eu_shifts}.} and are followed by a period of inactivity of at
least eight hours.\footnote{The US Department of Labor considers that a normal
shift is followed by ``at least an eight-hour rest''~\cite{labor_shifts} 
while the European Union's 2003/88/EC directive establishes a ``minimum 
rest period of 11 consecutive hours.''}

Next, we exclude all tweets not belonging to the dominant time frame.
We also exclude clusters that repeatedly have daily activity of more than ten hours,
as they most likely do not correspond to the user's work (since we assume that most jobs 
have eight-hour shifts). However, as sometimes 
people are required to work overtime, or stay at work longer than usual, 
we are flexible and only exclude clusters with more than 20\% of
their daily time frames exceeding the ten-hour threshold (i.e., one workday per week)
based on reported average overtime hours in the US~\cite{labor_stats}
and the European Union's limit for 48 hours per week.
Finally, we select the cluster with the largest number
of active weeks as the user's workplace.
It is important to note that our approach provides the first
adaptive approach that dynamically identifies shifts or common working hours
for each individual user, contrarily to previous approaches that followed 
a simplistic approach of considering fixed working hours for all users 
(e.g., ``09:00-17:00'').

An example of the tweeting activity of two users from both home
and work is given in Figure~\ref{fig:scaterplot}. Both users'
locations were correctly identified by \system.
For the top user, tweets from work fall in a well-defined 
time frame (08:00-16:00), in contrast to tweets posted from home,
which cover almost all times of day. The bottom user exhibits a more 
erratic behavior with different work shifts within a week,
highlighting the need for our dynamic approach that adapts to 
different patterns.

\subsection{Identifying Highly-Sensitive Places}
\label{sec:sensitive}

While identification of a user's home and workplace is a significant privacy
risk, our goal is to also explore the feasibility of uncovering personal user
information that may be considered even more sensitive. As such, we want to
identify other places a user has visited that could be used to infer such
sensitive information.
\system identifies a user's Potentially Sensitive
Clusters (PSCs) which are in close proximity to
highly-sensitive venues, and determines whether the user actually
visited these venues. To label a cluster as potentially
sensitive, we estimate the cluster's mid-point coordinates and
use Foursquare's~\cite{fsq_api} venue API for retrieving information about
the nearest venues. 

We consider venues that are within a 25 meter radius from the cluster's
mid-point coordinates; we set a more restrictive threshold compared to the key
location clustering process to avoid potential false positives due to the small
number of tweets per cluster and density of PSCs. In practice, if \system is
offered as an auditing tool to users, these thresholds can be user-configurable
to allow for flexibility for areas of different venue density (e.g., downtown
metropolitan areas vs rural areas).
The Foursquare API
returns the name of each venue as well as its type,
selected from an extensive list of predefined categories.
As such, we have identified which of the venues returned by the
API are associated with sensitive categories or subcategories (the categories 
we consider as sensitive in this study can be seen in Figure~\ref{fig:sensitive_clusters}).

\textbf{Content-based corroboration.} Proximity to a sensitive venue does not necessitate
that the user visited it (at least on that occasion). It 
could quite possibly be a case of simply passing by or visiting a different (potentially non-sensitive)
nearby venue. To determine if the user is associated with the sensitive 
venue, we analyze the content of the cluster's tweets in an effort to capture 
terms that indicate the user's presence at that venue. It is important to 
note that despite the user including some relevant keyword in the tweet,
location metadata allows attackers to \emph{obtain more context and 
infer sensitive information that the user did not intend to disclose}.

\system uses three manually-curated wordlists of related
terms based on numerous online domain-specific corpora
that contain keywords related to our sensitive categories.
Specifically, our wordlists contain medical- and health-related terms,
terms associated with various religions, and sex/nightlife.
We remove relevant keywords that are 
overtly ambiguous in context, as they can lead to false positives 
(e.g., ``joint'' may refer to a part of the body, some type of establishment, or may be drug-related).
Our wordlists are available online.\footnote{\url{https://www.cs.uic.edu/~location-inference/}}
These lists can be easily modified, or expanded 
to include terms from other categories (e.g., political).

\system first pre-processes users' tweets (i.e., tokenization, 
lemmatization, removes punctuation, emojis, mentions, stop-words and URLs) using the NLTK library.
Then it uses \textit{term frequency - inverse document frequency} (\texttt{tf-idf}) 
to identify the most significant terms within the tweets of each PSC.
For each cluster we consider the cluster's tweets as the \textit{document}
and the entirety of the user's tweets as the collection
(with each cluster considered a document).
As \texttt{tf-idf} assigns a score to the terms of the cluster, we check 
the three terms with the highest score against the respective wordlist, to
determine if the context of these terms can be associated with a nearby
sensitive venue.

\textbf{Duration-based corroboration.} Due to the sensitive nature of these
venues, users will not always include content in their tweets that enables us
to place them in a sensitive venue.
For this reason, we introduce another approach that does not depend on the content
of tweets, but on the repetitiveness and duration of user visits to a specific geographic
area, in order to identify places the user has likely visited. More specifically, with this
approach we identify PSCs that have consecutive tweets in the
span of a few hours, which indicate that the user has spent a considerable amount of time
at that place. In order to avoid cases where the users did not visit a sensitive place but
posted multiple tweets while passing by it, we exclude cases of consecutive tweets that
have been posted in short periods of time (within five minutes). We also identify
tweets posted from the same cluster on different days, which shows that the
user tends to repeatedly visit that place. Obviously this approach does not
work for clusters with a single tweet, 
and it lacks the additional confidence in placing the user at the sensitive venue 
that we obtain with the content-based approach. Nonetheless, it highlights a 
significant source of privacy leakage.

\subsection{Implementation Details}

\system has been designed as a completely modular framework, allowing for each
individual component to be trivially changed or extended (e.g.,
incorporating a new data source, or
implementing a different clustering method etc.). Our system has been
fully implemented in Python, and all collected data is stored into a Mongo
database.  In more detail, we leverage the \texttt{Tweepy} package for
interacting with Twitter's API and collecting users' timelines.  For the
first-level clustering and address validation we rely on the \texttt{Geopy}
package (via which we interact with the ArcGIS and Google APIs), while our
second-level clustering is based on the default implementation of DBSCAN as
provided by the \texttt{scikit-learn} package. For collecting venue information
\system uses the \texttt{Foursquare} package, while the \texttt{NLTK} package
is used for all tweet preprocessing and procedures related to tf-idf.  Given
the importance of scalability when processing large collections of users, we
have designed \system to be able to use multiple API keys in parallel.
This allows us to speed up the more inefficient parts of the process which rely
on communicating with external, and often rate-limited, APIs. Finally, as each user
is processed completely independently from other users at all stages, multiple
instances of our framework can be executed in parallel for increasing
efficiency.

\section{Data Collection}\label{sec:collection}

We first describe
our automatically-collected Twitter datasets, and then outline our methodology
for manually creating a ground truth dataset used for the experimental 
evaluation of \system in Section~\ref{sec:evaluation}.

\textbf{Datasets.}
We used Twitter's streaming API for collecting a set 
of tweets within a bounding box that covers the mainland 
area of the United States. While \system can be applied to any
country with similar working norms (e.g., shift duration)
we opted for users in the US as our sensitive location inference
also requires the tweet content and we currently only support English.
Furthermore, it is also the one country common across the datasets
of all the prior studies we compare to in Section~\ref{sec:evaluation}.
Nonetheless, an interesting future direction is to explore these 
privacy risks for users in other countries.

An initial set of tweets was collected 
in November 2016, through which we obtained 308,593 unique user identifiers (UIDs).
Then we collected each user's profile information and timeline
(the 3,200 most recent tweets, according to Twitter's policy).
This dataset contains 456,856,444 tweets, which have
been generated from 15,094 distinct sources (including unofficial Twitter
client apps and websites.) 

\begin{figure}[t]
	\centering
	\includegraphics[width=\columnwidth]{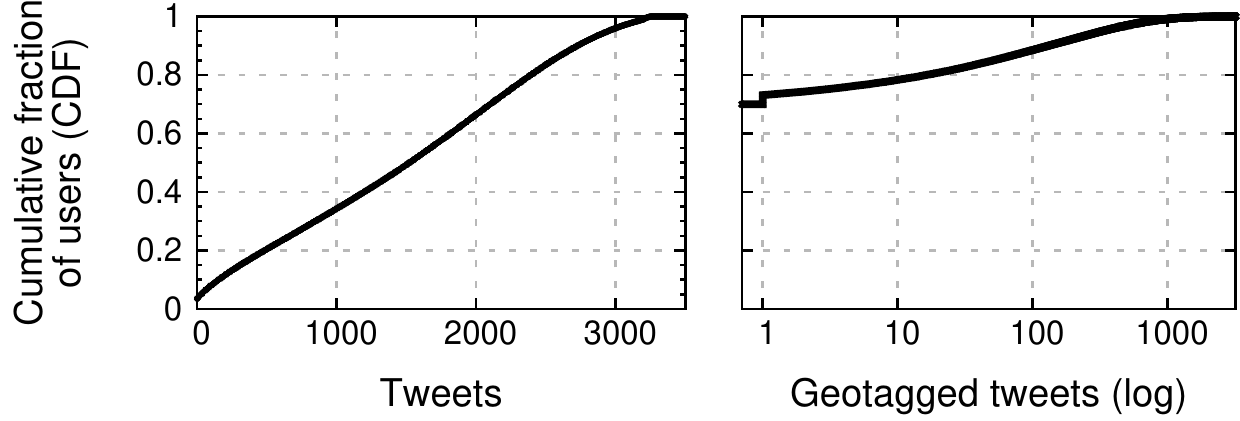}
	\caption{Total number of tweets per user (left), and the number of tweets per user that are geotagged (right).}
	\label{fig:cdf_users}
\end{figure}

Apps may 
handle geolocation data differently as Twitter's Geo Guidelines~\cite{twitter_guidelines} 
are neither mandatory nor enforceable.
To avoid inconsistencies, 
we only consider official Twitter apps and Foursquare
in this study, which 
also account for the vast majority of collected tweets.
After this filtering, we end up with 290,162 users
and 345,643,445 tweets.
We break down our dataset in Table~\ref{tab:sources};
users who posted tweets from multiple apps
are counted in all the respective categories.
Figure~\ref{fig:cdf_users} (left) shows the number of tweets in each
user's timeline. We find that 
only $\sim$0.5\% of the users have more than 3000 tweets, and
less than 0.06\% reached Twitter's API limit of 3,200.

As we are interested in the privacy 
implications that stem from geolocation 
metadata, we identify all users with
at least one tweet containing GPS coordinates in the metadata.
We identified 87,114 such users, which have 
contributed 15,263,317 geotagged tweets in total.
In Figure~\ref{fig:cdf_users} (right) we present
the number of users' geotagged tweets. Surprisingly we find that for 30.03\% of the users
the Twitter API reveals some precise geolocation information,
with 8.01\% of the users having less than 10 geotagged tweets.
We also observe that 
15.55\% of the users have between 10 and 250 geotagged tweets, 
and approximately 5\% and 2\% of the users have more than 330 and 
655 geotagged tweets, respectively.

Users with many geotagged tweets may have patterns 
that differ from those of users with a significantly lower number.
For this reason we conduct our analysis 
on two different sets of users. The first set (Top-6K) consists of the top 6,010
users in our dataset that have the most geotagged tweets (approximately top 2\% of
users in Figure~\ref{fig:cdf_users}), while the second set (Low-10K) 
consists of 9,841 randomly selected users that have between 10 and 250 
geotagged tweets. We use these two sets of users for our main analysis
(instead of all collected users), 
due to the rate limits imposed by the API providers that we use for our 
clustering process. Also, by including users with as few as 10 geotagged tweets,
we can explore the privacy risk that users face even when very few location 
data points are available.

\emph{Geotag accuracy.} It is important to note that while other types of
location-based services may add some form of noise or obfuscate the user's
location~\cite{polakis:ccs15}, that is not the case with Twitter. As such, 
the GPS coordinates contained in the tweet metadata we obtain through the 
API match those provided by the user's device.

\begin{table}
\caption{Breakdown of tweets' sources in our dataset.}
\centering
\small
\begin{center}
\begin{tabular}{l|c|c|c}
\hline
\textbf{Application} (source) & \textbf{Geoloc.} &\textbf{Users} &\textbf{Tweets}\\
\hline
\rowcolor{Gray}
\begin{tabular}[x]{@{}c@{}}Twitter for Android\end{tabular} &\cmark  & 99,979& 50,188,992\\ 
\begin{tabular}[x]{@{}c@{}}Twitter for iOS\end{tabular} &\cmark & 328,320 & 291,820,742\\ 
\rowcolor{Gray}
\begin{tabular}[x]{@{}c@{}}Twitter for Web\end{tabular} &\xmark  & 253,616 & 39,655,850\\ 
\begin{tabular}[x]{@{}c@{}}Foursquare\end{tabular} &\cmark  & 13,192 & 3,633,711\\
\hline
\end{tabular}
\end{center}
\label{tab:sources}
\end{table}

\textbf{Ground truth collection.} 
As we aim to demonstrate the true extent of this privacy issue by
identifying key locations at a postal address granularity, a significant
challenge is obtaining the ground truth for evaluating the accuracy of our
approach. While our home/work identification algorithms focus on
spatio-temporal characteristics, creating the ground truth
mandates an analysis of the tweets' content. Due to \emph{strict requirements
for veracity}, we did not resort to an automated process but opted for an arduous
and painstaking manual process that required over 6 weeks of continuous effort.
While we have explicitly limited our data collection to \emph{publicly
available data} offered by the official Twitter API, we took extra precautions
during our manual data analysis phase for protecting users' privacy. Specifically, the user's account
information (name, username) was not included in the content that was manually
inspected, and references to other users (i.e., tokens starting with ``@'')
were removed as well.

In a nutshell, we started with users 
that explicitly mentioned in their tweets that they are at home, by matching 
phrases such as ``Im home'', ``I'm home'', ``at home''
etc. After identifying users with clusters containing such tweets, we 
manually reviewed all the tweets in these particular clusters. During
the manual inspection we took into account the context of a user's 
tweets for ensuring that these clusters indeed correspond to a user's 
home.
Instead of identifying work clusters for other users, we decided to focus this
task on the users for which we have already identified their home location, as
that would allow us to create a more complete dataset that contains both home
and work locations for each user. To that end, we followed a similar approach
and searched for phrases denoting work-related information, ``at work'', ``at
the office'', ``my job'', ``this job'' etc., and manually inspected the tweets
of the returned clusters.

Below we outline the workflow of our manual inspection process for 
identifying users' home and work locations. Our goal was to establish a 
methodology
that allows us to have high confidence in the resulting labels. The content 
analysis and location labeling was performed by two researchers independently;
\emph{in cases where the labels by the two researchers did not match the user
was discarded.} We avoided potentially ambiguous instances or cases with uncertainty,
and built our ground truth with users where both labellers agreed.
We discarded such instances as we set a strict requirement for correct labels for our 
ground truth. However, discarding users was a rare occurrence, as it is a fairly 
straightforward and intuitive process for human annotators to identify home/work locations.
In more detail, we established the following workflow:

\begin{enumerate} 
\item Apart from inspecting the tweets that contained one of the initial seed phrases,
we also inspected all of the cluster's remaining tweets. This allowed us to 
further increase our confidence by identifying tweets where the user
explicitly or implicitly referred to being at home or work  (e.g., ``just took a shower'', 
``my boss just said'' etc.). If we only found implicit references, we 
required at least two such tweets to increase our confidence.

\item 
To make our ground truth as complete as 
possible, we also manually inspected all the tweets in users' 10 largest
clusters, for identifying cases where users have multiple homes or
work clusters that were not already identified during the previous task.
Again we followed the same approach as described in the previous step.
In cases where there were no other clusters with tweets indicating a home or work location
we were confident of our original labeling, since there was only one cluster matching 
each label. In cases where other clusters'
alluded to a potential key location, we continued with the following process:

\begin{enumerate} 
\item \emph{Temporal analysis.} 
We explicitly analyzed the timeline of clusters, and identified the periods
during which each cluster was active. This helped us identify cases
where users had changed residences, where multiple locations had been
labeled as homes but their active periods did not overlap temporally.
We also observed cases where the identified home
was not the user's place of residence, but could be
considered a secondary home (i.e., country/summer house, 
parents' house). During this step we also searched specifically for
references that allowed us to label the cluster as a secondary home location
(e.g., terms referring to parents).

\item \emph{Spatial analysis.}
In cases where more than one cluster exhibited home-like patterns
and had overlapping active periods, we considered the spatial location 
of each cluster.
If the two clusters were close geographically, we further investigated them to decide
which one was the user's actual home and which was not (e.g., a
friend's house that the user visits frequently). For clusters that were far
away from each other (e.g., in different cities), we relied on the content 
for verification.
A common occurrence was clusters with home-related keywords that exhibited 
continuous activity for a few days: e.g., users tweeting that they were at home,
while visiting their parents' house during the holidays.
\end{enumerate}

\end{enumerate}

Overall, in the Home-Top dataset we have 1,004 users with 1,307 
home clusters; 718 of these users have only one 
home cluster, while 269 and 17 users have two and three homes, 
respectively. This is not a surprising finding, as we collected all the tweets in each 
user's timeline (up to 3,200), and not only tweets posted in a specific time period. Indicatively, 
we have observed cases of users that have relocated (e.g., after graduating),
college students living in dorm rooms during their first year
and then moving to a house, and students that regularly 
visit their family home. We also observed users with multiple 
home locations in the Home-Low ground truth dataset,
but to a lower extent. Specifically, we identified 905 users that have
one home cluster, 137 users with two, and one user with three home clusters.
For the two work ground truth datasets, i.e., Work-Top and Work-Low, we 
identified 298 and 92 users, that have 363 and 98 work clusters respectively.

It is possible that our home/work ground truth datasets are not
exhaustive (i.e., we may have missed certain locations).
However, due to the systematic and stringent manual inspection process, we are certain
that all the locations labeled in our ground truth
indeed correspond to users' home and work locations. Our manual inspection 
process has resulted in ground truth datasets significantly more complete and fine-grained 
than those used in prior studies~\cite{efstathiades_2015,hu_2015,cheng2011exploring,cho_2011,lin_2017}
(more details can be found in Section~\ref{sec:related}).

\begin{figure}[t]
	\centering
	\includegraphics[width=\columnwidth]{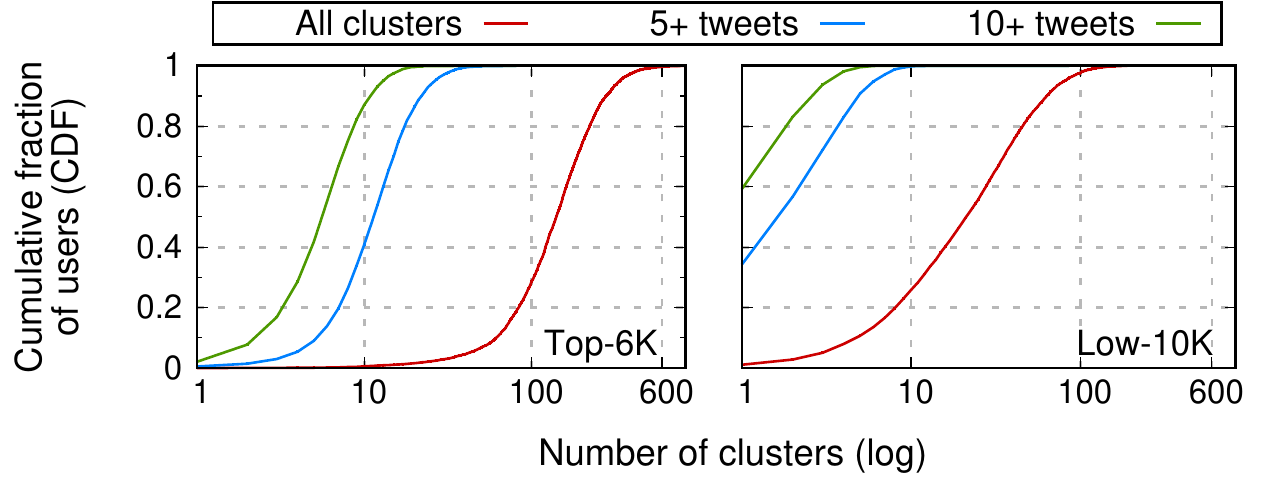}
	\caption{Number of clusters per user.}
	\label{fig:cdf_num_clusters}
\end{figure}

\section{Experimental Evaluation}
\label{sec:evaluation}

In this section we first present an analysis of our datasets
and discuss properties of users' behavior regarding key and sensitive locations.
Subsequently, we use our ground truth to experimentally evaluate \system and compare to prior work.

\begin{figure*}[t]
	\centering
	\begin{minipage}{0.31\textwidth}
		\centering
		\includegraphics[width=\textwidth]{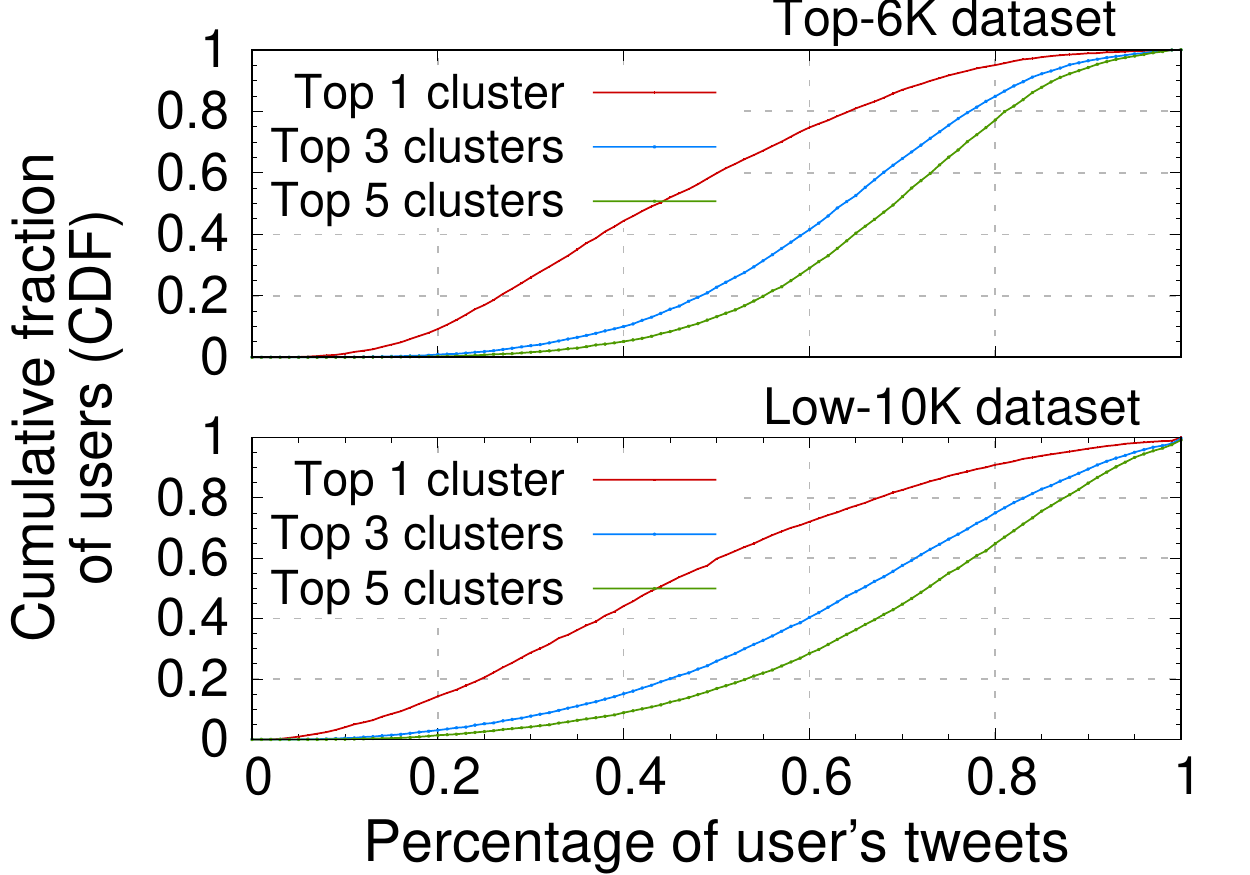}
		\caption{User's tweets from their top 1, 3, and 5 clusters.}
		\label{fig:cdf_top_clusters}
	\end{minipage}
	\hfill
	\begin{minipage}{0.31\textwidth}
		\includegraphics[width=\textwidth]{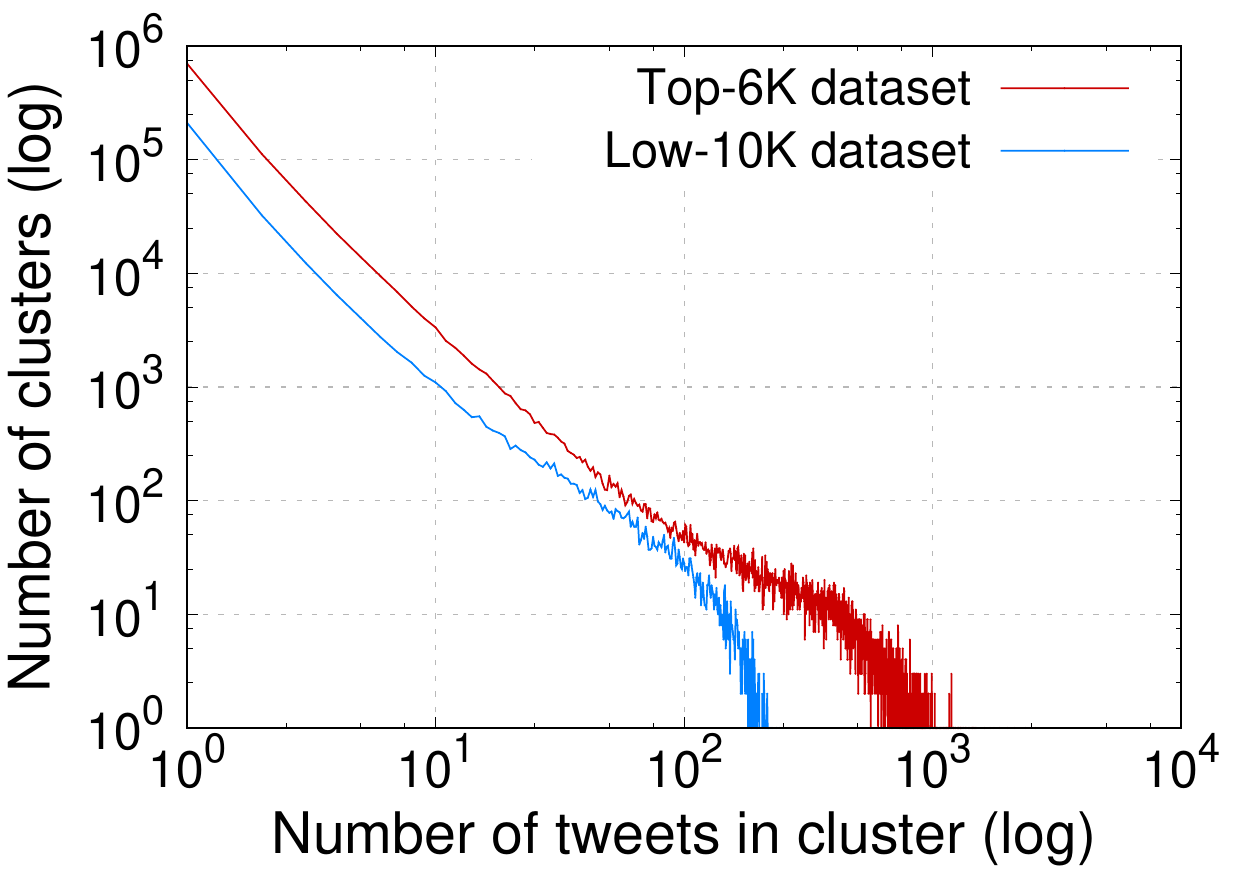}
		\caption{Tweets per cluster in the two datasets.}
		\label{fig:cluster_size}
	\end{minipage}
	\hfill
	\begin{minipage}{0.31\textwidth}
		\includegraphics[width=\textwidth]{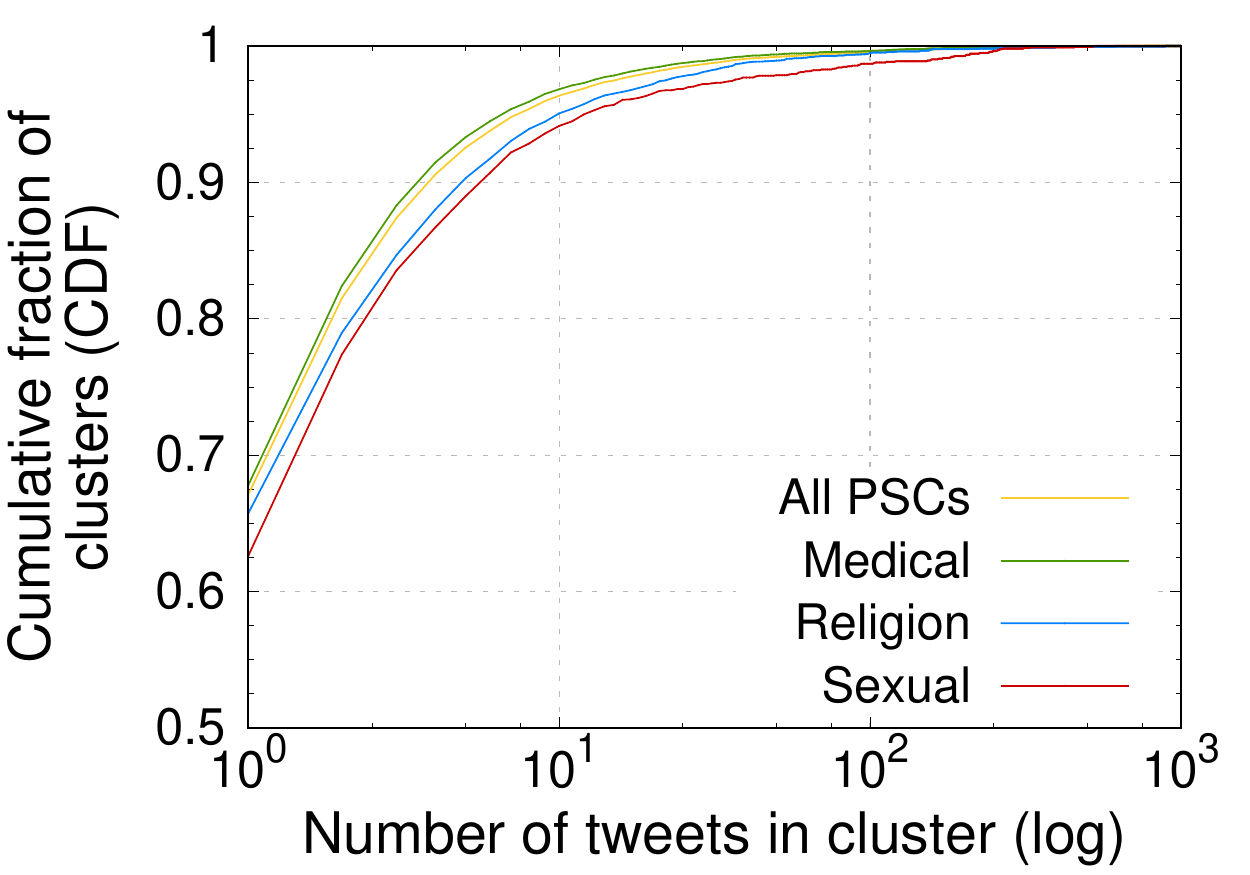}
		\caption{Tweets from Potentially Sensitive Clusters (PSCs).}
		\label{fig:cdf_PSCs}
	\end{minipage}
\end{figure*}

\textbf{Location clusters.} 
To investigate the location patterns in users' tweeting behavior,
we focus our analysis on understanding the characteristics of users'
location clusters. We perform this analysis for both the most active 
(Top-6K) and less active (Low-10K) users.
Figure~\ref{fig:cdf_num_clusters} depicts the number of clusters per user.
As expected, highly active users tend to have a
large number of clusters. Specifically, only 4.45\% of these users 
have less than 40 location clusters, and around 28\% less than 100 
clusters. In more detail, we observe that around 50\% 
of the highly active users have more than 140 clusters, and 
about 25\% and 10\% of them have more than 200 and 280 clusters
respectively. If we only consider 
clusters that have more than five tweets, we observe that about 50\% 
of the users have more than 11 such clusters, and 
10\% have more than 22 clusters. 

When focusing our analysis on the Low-10K dataset, we observe that 
these users have significantly less clusters than the 
highly active users but seem to follow a 
similar pattern. As shown in Figure~\ref{fig:cdf_num_clusters} 
(right), about 10.7\% have five or less clusters, 
and about 50\%, 25\% and 10\% of the users have more than 
21, 40 and 63 clusters respectively. Furthermore, similarly to the 
highly active users, the number of clusters drops significantly
when considering only those clusters that have more than 5
tweets. For both sets of users we find that users 
tend to have a large number of clusters, out of which the 
majority has a small number of tweets. 

Figure~\ref{fig:cdf_top_clusters}
presents the percentage of users' tweets in their five largest clusters.
We observe that for about 40\% of the users, 
more than half of their tweets belong to their top cluster, while 47.77\% of the 
users have more than 70\% of their tweets in their top 5 clusters.
This phenomenon is observed in both sets of users.
In Figure~\ref{fig:cluster_size} we explore the cluster sizes of all users.
Both datasets exhibit a power law distribution, with 
the vast majority of clusters having only a few tweets and a small 
number of clusters with a large number of tweets.
These small clusters will most likely not correspond to a user's \textit{home} 
and \textit{work} locations, as they appear to be visited rarely;
however, these locations are important from a privacy perspective, 
as they allow an adversary to reconstruct a semantically-rich location 
history, which can reveal highly sensitive information.
In fact, this is clearly demonstrated in Figure~\ref{fig:cdf_PSCs}, which
presents the distribution of PSCs with 
regards to the number of their tweets.
We find that 67.10\% of the PSCs in our datasets have
one tweet, while only 4.04\% of them have 10 or more (the most being health-related).

\subsection{Home and Work Location Inference}

To assess our methodology and measure the effectiveness
of \system
we aim to pinpoint exact locations.
Thus, we opt for a ``strict'' evaluation of accuracy where a location is either
correctly or incorrectly identified. We do not calculate
distance errors as they are more suitable for coarse-grained 
approaches that roughly estimate locations.

\begin{table}
	\caption{Performance of home/work inference for ground truth users,
		and ranks of the respective clusters.}
		\label{tab:home_work_detection}
	\centering
	\small
	\begin{center}
		\resizebox{\columnwidth}{!}{
		\begin{tabular}{cccc|ccccc}
\toprule
			&  &
			\multirow{2}{1.3cm}{\centering \textbf{Inferred clusters}} &
			& \multicolumn{5}{c}{\textbf{Rank of clusters}}\\
			\textbf{Dataset}&\textbf{Users} &  & \textbf{Precis.} &1 & 2 &3 & 4 & 5-10\\
			\midrule
\rowcolor{Gray}
Home-Top & 1004 & 926 & 92.2\% & 806 &111 & 8 & 1 & - \\ 
			Home-Low & 1043 & 969 & 92.9\% & 911 & 49 & 8 & - & 1 \\
			\midrule
\rowcolor{Gray}
			Work-Top & 298 & 164 & 55\% & 7 & 79 & 47 & 16 & 15 \\
			Work-Low & 92 & 53 & 57.6\% & 4 & 31 & 11 & 6 & 1 \\
\bottomrule
		\end{tabular}
}
	\end{center}
\end{table}

\system correctly identifies the home of 926 and 969 users
from the two datasets, resulting in a precision of 92.23\% and 92.9\% respectively.
Thus apart from obtaining superior granularity, our system
is considerably more effective than previous approaches as we will show.
As our work inference first excludes the home cluster, the outcome also depends
on the precision of the home inference. Our precision is 55.03\% and 57.6\%
for identifying workplaces in our ground truth.  As users typically tweet less
when they are at work than when they are at home (in our ground truth, home
clusters contain an average of 45\% of tweets while work clusters contain 8\%) our
effectiveness at identifying work is lower since other locations frequented by
the user can exhibit similar characteristics (e.g., restaurants, coffee shops,
gyms).  Table~\ref{tab:home_work_detection} presents the precision of our home
and work inference, as well as the rank of all the correctly identified
clusters.  The clusters' ranks are estimated according to their size, such that
rank 1 is the largest cluster of the user, rank 2 is the second largest cluster
and so on. Also, we do not re-calculate cluster ranks after excluding home clusters
in the work identification phase, as we want to make direct comparisons between
the results of the two approaches. 
Finally home clusters have, on average, a maximum radius of 59.55 meters and work clusters of 53.38,
which drops to 19.25 meters for all clusters in our ground truth.

\begin{figure}[t]
	\centering
	\includegraphics[width=0.85\columnwidth]{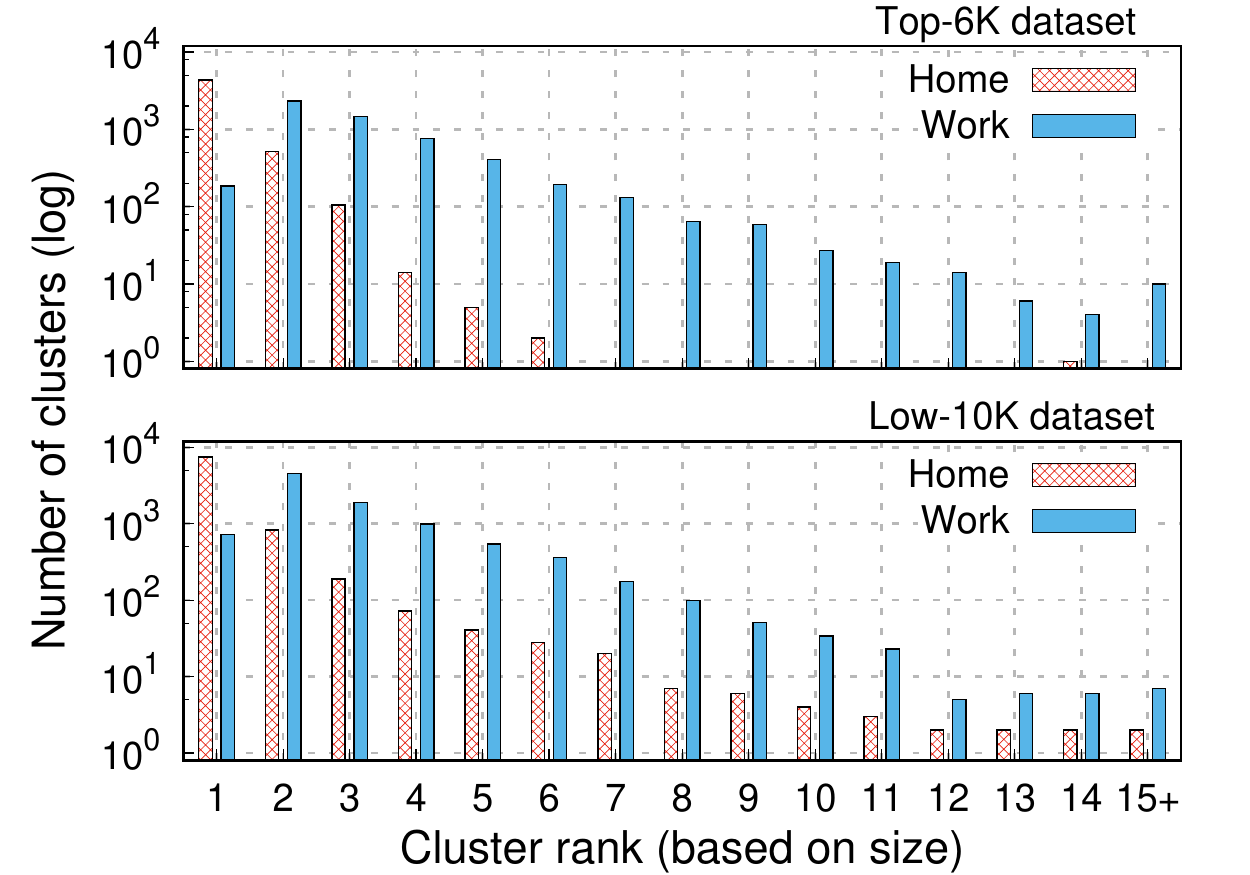}
	\caption{Ranks of home and work clusters for our main datasets (ground truth users have been excluded).}
	\label{fig:large_dataset_ranks}
\end{figure}

Having established the precision of \system 
on our ground truth datasets, 
we run our system on the main datasets
(Top-6K, Low-10K) after excluding the ground truth users.
As can be seen in Figure~\ref{fig:large_dataset_ranks}, the majority of 
home clusters in both 
datasets are rank 1 clusters, which is consistent 
with the results from the home ground truth.
For the work clusters, only 3.26\% and 7.69\% are rank 1, while most of them 
are rank 2
and a considerable number occupy lower ranks, in both datasets.
We find that the detected clusters follow a similar rank distribution in the
two datasets, supporting the representativeness of our ground truth.  
It should be noted though that while our work ground truth
explicitly contains users for which we have identified their work, for the main
datasets our system also identifies locations that are not work in the
strictest sense.
Specifically, we are able to identify locations for users
that do not work but have a location that can be considered a work
``substitute'',
e.g., a college student attending classes.

\textbf{Selection bias.} The methodology that we employed to create our ground
truth datasets could potentially result in selection bias, as it relies on certain key phrases
as a starting point for the manual process. To examine whether the accuracy of our evaluation is a 
byproduct of \system's heuristics being ``overfitted'' to the ground truth, we manually examine
a random subset of users identified from the main datasets (Top-6K, Low-10K). Specifically, we 
select 100 users and manually investigate their tweets to verify whether the home and work labels
assigned by our system are correct. Following the manual methodology described in Section~\ref{sec:methodology}
we are able to verify that 89 of the home labels indeed correspond to the user's actual home cluster.
For the remaining 11 users we are unable to characterize the label as correct or incorrect based
on the users' tweets. For the work labels, we find that for 45 users the work cluster has been 
correctly identified while for 30 users the label is incorrect. For the remaining 25 users we 
are not able to verify whether the label is correct or not.

While this manually verified sample is relatively small, we find that the
resulting accuracy is comparable to the accuracy achieved by our system when
evaluated against the ground truth. Furthermore, these users are from
our main datasets, which exhibit a wide range of geotagging behavior,
demonstrating that our effectiveness is not tied to a
specific dataset. As \system's underlying algorithms are based on common user
behaviors and legislative/societal norms, we believe that this manual
verification further validates the generalisability of our techniques and the
correctness of our ground truth.

\textbf{De-anonymization.} While demonstrating the feasibility of
de-anonymizing Twitter users is not the focus of
our work, we conduct a small exploratory experiment. We aim to identify which,
if any, users in our ground truth datasets appear to be pseudonymous.
Specifically we want to identify users that do not provide their full name,
i.e., do not provide their last name (we do not consider first names to
be conclusive for identity). We use the list provided by
the US Census Bureau with the most frequent surnames to filter 
out users that include their last name in the \textit{full name} section of their account.

After filtering 282 users remain, which we manually examine and exclude the
ones that actually disclose a last name that is not included in the Census
list, or include their last name in their username. We end up with 183 users
that do not explicitly reveal their identity on their Twitter accounts, which
constitutes a lower bound of the pseudonymous users in our ground truth, due to
potential false positives in our automated filtering. Out of these users,
\system was able to correctly identify the home location of 171 users and the
workplace of 23 users (to ensure privacy, the manual inspection of users' names
was conducted in ``isolation'' and not combined with or mapped to any ground truth
locations or other location clusters). While these users might not be
\emph{truly} pseudonymous in reality (e.g., users with a pseudonym whose actual
identity is well known within certain communities) or could also be
de-anonymized through other techniques~\cite{Goga:2013}, this
experiment highlights another potential threat posed by location metadata.

\begin{table*}
	\caption{Comparison between the precision achieved by \system
    and previously proposed approaches. We have implemented 
    all prior heuristics and applied them to our ground truth datasets 
    to allow a direct comparison.}
	\label{tab:comparison}
	\centering
	\small
	\begin{center}
		\begin{tabular}{cc|l|c|c|c}
            \hline
			& & \multirow{2}{9.5cm}{\centering \textbf{Heuristic Description}} &
			\multicolumn{2}{c|}{\textbf{Dataset}} &
			\multirow{2}{2.1cm}{\centering \textbf{Proposed by}} \\
			
			& & & \textbf{Top} & \textbf{Low} & \\
			\hline
			
			\multirow{13}{*}{\centering \textbf{Home}} & 1 & 
			Cluster with the highest number of tweets
			& 72.3\% & 67.8\% & \cite{cheng2011exploring, cho_2011, hu_2015, krumm2007}\\ \cline{2-6}

			& 2 & Most tweets between 20:00-8:00 
			& 72.1\% & 66.4\% & \cite{Luo:2016}\\ \cline{2-6}
			
			& 3 & Most tweets between 24:00-7:00 
			& 69.3\% & 54.7\% & \cite{hu_2015}\\ \cline{2-6}

			& 4 & Last destination of the day (before 3am) 
			& 73.3\% & 64.8\% & \cite{hu_2015, krumm2007}\\ \cline{2-6}

			& 5 & Last destination of the day (w/o days with tweets between 24:00-7:00)
			& 71.4\% & 64.4\% & \cite{hu_2015} \\ \cline{2-6}
			
			& 6 & Weighted PageRank for destinations
			& 44.1\% & 26.4\% & \cite{hu_2015}\\ \cline{2-6}

			& 7 & Weighted PageRank for origins
			& 37.5\% & 20.9\% & \cite{hu_2015}\\ \cline{2-6}

			& \multirow{2}{*}{8} & \multirow{2}{9.5cm}{Most popular cluster in terms of unique days, during the \textit{Rest} (2:00-7:59) and \textit{Leisure} (19:00-01:59) time frames} 
			& \multirow{2}{*}{73.1\%} & \multirow{2}{*}{64.9\%} & \multirow{2}{*}{\cite{efstathiades_2015}}\\
			&&&&&\\ \cline{2-6}

			& 9 & WMFV (best reported time frame: 24:00-5:59)
			& 65\% & 50.9\% & \cite{lin_2017}\\ \cline{2-6}

			& 10 & W-MEAN (best reported time frame: 24:00-5:59)
			& 0.6\% & 14.7\% & \cite{lin_2017}\\ \cline{2-6}

			& 11 & W-MEDIAN (best reported time frame: 23:00-5:59)
			& 15.6\% & 24.5\% & \cite{lin_2017}\\ \cline{2-6}

			& 12 & \system's Home detection without 2\textsuperscript{nd} level clustering
			& 73.7\% & 69.3\% & this paper\\ \cline{2-6}

			& 13 & \textbf{\system's Home detection}
            & \textbf{92.2\%} & \textbf{92.9\%} & \textbf{this paper}\\

			\midrule
			
			\multirow{4}{*}{\centering \textbf{Work}} & \multirow{2}{*}{14} & \multirow{2}{9.5cm}{Most popular cluster in terms of unique days, during the \textit{Active} time frame (e.g., working hours, 08:00-18:59)} 
			& \multirow{2}{*}{33.2\%} & \multirow{2}{*}{48.9\%} & \multirow{2}{*}{\cite{efstathiades_2015}}\\
            &&&&&\\ \cline{2-6}
            & 15 & Cluster with the second highest number of tweets & 18.5\% & 22.8\% & - \\
            \cline{2-6}
            & 16 & \system's Work detection without 2\textsuperscript{nd} level clustering & 32.2\% & 30.4\% & this paper \\
            \cline{2-6}

			& 17 & \textbf{\system's Work detection}
            & \textbf{55\%} & \textbf{57.6\%} & \textbf{this paper}\\
            \hline

		\end{tabular}
	\end{center}
\end{table*}

\textbf{Prior work.}
Apart from pinpointing locations
with a granularity that is orders of magnitude more fine-grained than prior work, it is important
to also quantify the accuracy improvements of our techniques. We 
implement the heuristics proposed in prior work for identifying home and work 
locations that leverage spatiotemporal patterns and apply them to our ground truth;
we do not compare to techniques that require other types of data, like social ties~\cite{Huang:2016,Huang:2017},
as we do not collect such data and those techniques are inherently very coarse-grained.
By running these heuristics on the same data, we are able to conduct a
direct comparison to previous techniques
and avoid the inaccuracy of simply comparing to their reported numbers.
It is important to note that we map tweets to postal addresses before
applying these previously-proposed heuristics, i.e., we only apply our initial first-level clustering 
so as to remain as faithful as possible to their original design.

As Table~\ref{tab:comparison} shows, \system outperforms all
heuristics proposed in prior work for both home and work locations. The
simplistic approach of selecting the largest cluster as the home (1)
performs surprisingly well, and even outperforms some of the other more complex
heuristics.
We also extended this logic and evaluated the precision of
considering the second largest cluster as the workplace (14); this results in a
precision of 18.45\% and 22.82\% in the Work-Top and Work-Low datasets respectively.
Heuristics (4) and (8)
perform better than other prior heuristics.
The approaches proposed in~\cite{lin_2017} rely on weights obtained from their data;
to remain faithful to their design, we replicate their approach
and randomly select 22\% of our users as the \textit{sample dataset} to calculate the weights
and the rest as the evaluation dataset.
The significant difference between their reported accuracy and our findings
can be attributed to their experiments being conducted on a dataset from a very limited 
time frame and geographic area.

Overall, our techniques present an improvement of 18.9\%-91.6\% when 
inferring homes and 
8.7\%-21.8\% for workplaces.
An interesting observation is that in multiple cases \system presents a larger
improvement over prior approaches for users that are not prolific geotaggers (i.e., from the Low datasets),
indicating the benefit of our techniques when there is sparser availability of data.

In an effort to accurately quantify the effect of our second-level clustering on our
techniques, we also run our heurstics by applying only the first-level clustering and compare
them to prior work. LPAuditor's home inference still outperforms all previous approaches, both
in the Home-Top and Home-Low datasets, with an improvement of 0.4\%-73.1\%.
On the contrary, heuristic (14) outperforms LPAuditor's work inference in both Work-Top and
Work-Low datasets, by 1\% and 18.5\% respectively.
These differences in our results can be attributed to the fact that when a user's location
cluster is split into smaller clusters (i.e. first-level clusters), our heuristics cannot entirely
capture the true behavior of that location, which further signifies the importance of our
second-level clustering. As aforementioned, other locations frequented by users during or near 
working hours can exhibit work-like characteristics (e.g., gym, coffee shopts etc.) affecting
fine-grained approaches like ours when working with sporadic location datasets. 
Our second-level clustering allows our system to group data points that belong to the same 
location but have been assigned to nearby locations due to the displacement introduced by user
mobility (e.g., tweeting while leaving work) or GPS errors. In a sense, this can be seen as enhancing 
the ``signal'', thus allowing our system to better capture the user's behavior
in each location. After applying our second-level clustering, our techniques improve significantly by
18.5\% and 23.6\% for the Home-Top and Home-Low datasets and by
22.8\% and 27.2\% for the Work-Top and Work-Low datasets respectively.

\subsection{Inference of Sensitive Places}

\begin{figure}
	\centering
	\includegraphics[width=1\columnwidth]{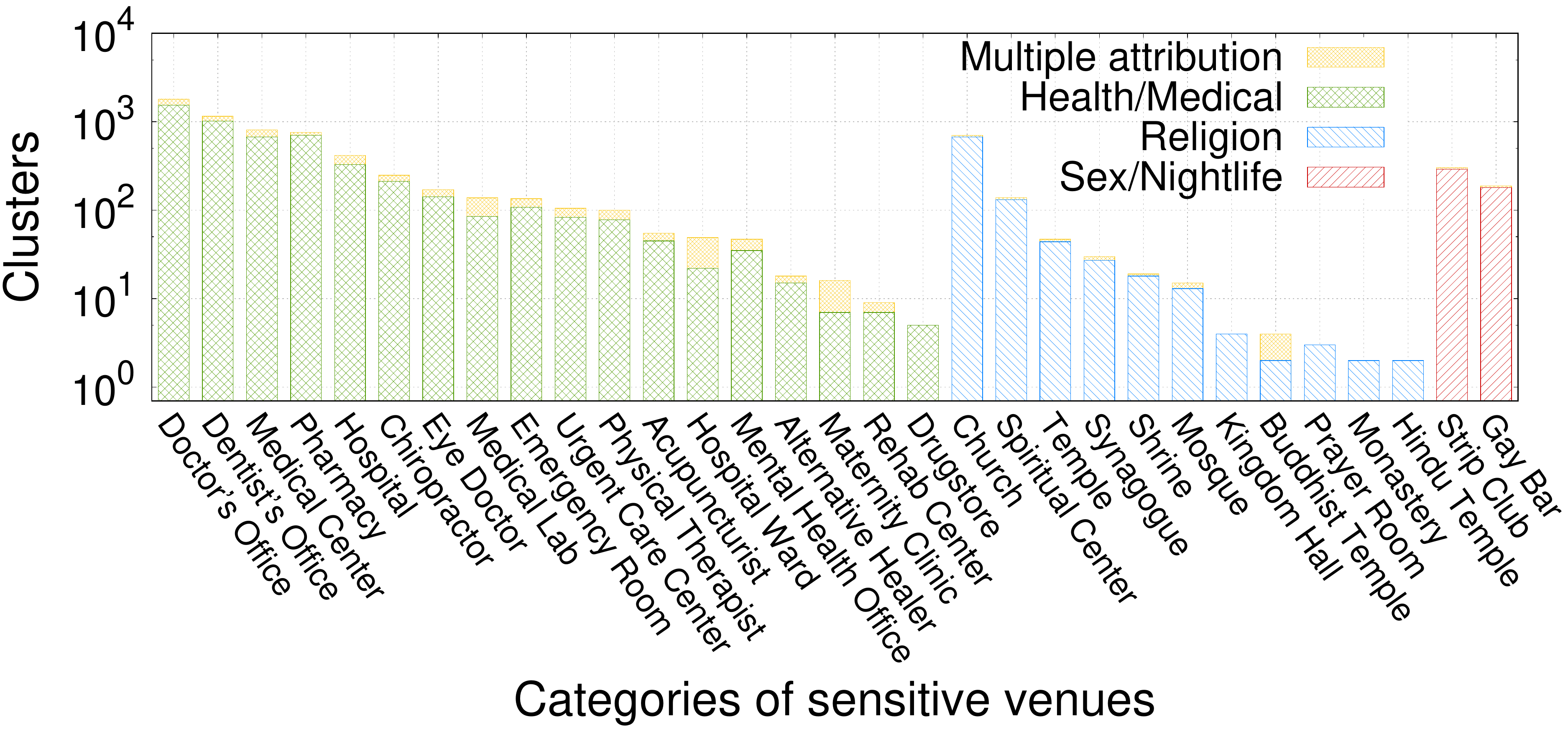}
	\caption{Potentially sensitive clusters, i.e., in close 
		proximity to venues belonging to a sensitive category.}
	\label{fig:sensitive_clusters}
\end{figure}

\system detected 6,483 potentially sensitive clusters (PSCs) 
across our ground truth. Specifically, it identified 938 (93.42\%) Home-Top 
users with a total of 5,393 PSCs, and 516 (49.47\%) users in Home-Low
with 1,090 PSCs. This difference between datasets is
expected as users in the latter have fewer geotagged
tweets and considerably less clusters.
Figure~\ref{fig:sensitive_clusters} breaks down the detected PSCs
according to the category of the associated venues. For PSCs that have more 
than one sensitive venue in close proximity, we first assign that PSC to the 
category of the closest venue. We also present how the distribution
changes if each PSC is mapped to all sensitive venues in proximity (denoted as 
``Multiple attribution''). When only considering the
sensitive venue with the shortest distance to the PSC's coordinates, we 
identify 5,094 health-related clusters, and 918 and 471 
venues related to religion and sex/nightlife respectively.
Interestingly, if we intersect these clusters with users' ground truth work locations
we find 10 common instances in the Work-Low and 15 in the Work-Top sets;
out of those only 3 from the latter set were identified by our system through
\texttt{tf-idf}. As such, we believe that the vast majority of cases are users
visiting these sensitive venues, as opposed to working there.

\textbf{Content-based corroboration.} When using \texttt{tf-idf} and our
wordlists, we increase our confidence in placing users at 545
of the detected PSCs.
To assess these results we identified the clusters 
that contain at least one keyword from the respective wordlists and manually
inspected the clusters' tweets, to assert whether the user was 
actually referring to a sensitive place. This manual inspection showed that
our approach had an overall precision of 80.36\% and a 93.79\% recall, as 
presented in Table~\ref{tab:venues_detection}. Out of the 438 verified sensitive venues,
375 were related to health, 51 and 12 to religion and sex respectively.
We observed a small number of false positives due to ambiguous keywords
that remained in our wordlists (e.g., ``shot''); however, we kept these terms
as the true positives significantly outweighed the false positives.
On the other hand, in some cases our approach missed certain sensitive clusters due to users
that post sensitive content repeatedly from many clusters (e.g., a religious user
that tweeted religious content from multiple locations),
which resulted in these keywords not being deemed significant by \texttt{tf-idf}.
Furthermore, it is important to stress that we obtain a lower bound on the number 
of sensitive venues that a user has visited, as the user may simply post tweets that 
do not contain the appropriate context.

Depending on the attacker's end goal, there might not be a need for absolute certainty
of whether the user visited the sensitive place. Even low confidence levels may 
be considered a sufficient indicator; for instance, an insurance company looking at 
a user's social media profile to decide on adjusting the user's premium or purchasing 
their policy~\cite{insurance}. Nonetheless, as an extra source of ground truth,
we identified these users' tweets that 
were generated by the Foursquare app and followed the typical format of a check-in;
we then compared the venues of these check-ins
to the clusters of sensitive nearby venues. This allowed us to verify certain detected sensitive 
places irrespective of the content posted from these clusters. This
returned 105 sensitive clusters for our ground truth users, 20 of which were 
also detected by \texttt{tf-idf}. While this source of ground truth is considerably 
small, it offers an interesting indication of user behavior; users are extremely reserved 
when it comes to explicitly publishing that they are at a sensitive location.
This further exemplifies the implications of the location metadata being exposed,
as it directly undermines privacy-conscious user behavior.

To further investigate the tweeting behavior of users from sensitive 
venues, we used \system to infer sensitive places visited by the remaining users
from the Top-6K and Low-10K datasets. Our system identified 21,863 PSCs for 4,418
users from the Top-6K dataset and through content-based corroboration identified 
1,512 of them as sensitive clusters that have been visited. Of those, 1,282 are health 
related, 196 pertain to religion and 34 are 
related to sex. Similarly for the users from the Low-10K dataset, we 
identified 6,918 PSCs, with 474 being flagged by our system,
with 341 related to health, 115 to religion, and 18 to sex.

\begin{table}
    \caption{Results of content-based (CB) identification of users visiting sensitive places, for our ground truth.}
	\label{tab:venues_detection}
	\centering
	\small
	\begin{center}
		\resizebox{\columnwidth}{!}{
		\begin{tabular}{lccc}
            \toprule
			& \textbf{Home-Top} & \textbf{Home-Low} & \textbf{Total}\\
			\midrule
            \rowcolor{Gray}
			Users in Dataset & 1,004 & 1,043 & 2,047 \\
			PSCs & 5,393 & 1,090 & 6,483 \\
            \rowcolor{Gray}
			Users w/ PSCs  & 938 & 516 & 1,454 \\
			\midrule
			Guessed Clusters (CB) & 464 & 81 & 545\\
            \rowcolor{Gray}
			Users w/ CB Clusters & 328 & 72 & 400 \\
			True Positive (TP) & 368 & 70 & 438\\
            \rowcolor{Gray}
			False Positive (FP) & 96 & 11 & 107 \\
			False Negative (FN) & 25 & 4 & 29 \\
			\midrule
            \rowcolor{Gray}
			Precision (TP/TP+FP) & 79.31\% & 86.41\% & 80.36\% \\
			Recall (TP/TP+FN) & 93.63\% & 94.59\% & 93.79\% \\
            \rowcolor{Gray}
			F-Score & 85.87\% & 90.31\% & 86.55\% \\
            \bottomrule
		\end{tabular}
}
	\end{center}
\end{table}

\textbf{Duration-based corroboration.} 
When using the duration-based approach (DB), as can be seen in 
Table~\ref{tab:passersby}, we identified 691 users from the 
Home-Top and 205 from the Home-Low dataset 
that have repeatedly visited or spent a considerable amount
of time at 1,699 and 276 PSCs respectively.
Similarly, in the Top-6K and Low-10K
datasets, we identified 3,012 and 1,672 users that have visited 7,020 and 
2,337 such places. It should be noted though that these numbers 
constitute a lower-bound estimation, as the duration-based 
approach does not take into consideration PSCs that only contain a single tweet.

\begin{table}
    \caption{Results of duration-based (DB) identification of users visiting sensitive places, for all datasets.}
	\label{tab:passersby}
	\centering
	\small
	\begin{center}
		\resizebox{\columnwidth}{!}{
			\begin{tabular}{lcccc}
                \toprule
				& \textbf{Home-Top} & \textbf{Home-Low} & \textbf{Top-6K} & \textbf{Low-10K}\\
				\midrule
                \rowcolor{Gray}
				Visited Clusters (DB) & 1,699 & 276 & 7,020 & 2,337 \\
				\quad \tabitem Medical & 1,307 & 194 & 5,193 & 1,626 \\
                \rowcolor{Gray}
				\quad \tabitem Religion & 245 & 56 & 1,176 & 493 \\
				\quad \tabitem Sex/nightlife & 147 & 26 & 651 & 218 \\
                \rowcolor{Gray}
				Users w/ DB Clusters & 691 & 205 & 3,012 & 1,672 \\
				\midrule
				Common CB/DB Clusters  & 53.44\% & 44.44\% & 53.9\% & 47.25\% \\
                \rowcolor{Gray}
				Users w/ CB/DB Clusters & 86.89\% & 59.72\% & 86.26\% & 65.88\% \\
                \bottomrule
			\end{tabular}
		}
	\end{center}
\end{table}

Interestingly, we observe that 53.44\% and 53.9\% of the sensitive 
clusters detected by the content-based approach (i.e., CB clusters) for 
the Home-Top and Top-6K datasets respectively, are among the visited 
clusters returned by the duration-based approach (DB clusters). For 
the Home-Low and Low-10K datasets, 44.44\% and 47.25\% of the 
clusters detected with the content-based approach have been also
detected by the duration-based approach. Employing both
approaches can increase confidence in identifying sensitive 
places the users have visited. 
Thus, for scenarios requiring higher levels of confidence, an attacker 
can select the intersection of the sets returned by the two approaches.
Another noteworthy observation is that the DB approach results in a higher 
ratio of sex-related clusters compared to CB, which indicates that users are 
reluctant to explicitly mention such venues while further highlighting the risk of
geotagged tweets.

\textbf{Contextual privacy loss.} A significant implication of this inference is 
that location metadata can amplify the loss of privacy by revealing sensitive details 
or additional context about the tweet's content that might not match the user's intended level
of disclosure. While we found this to be common across most cases of sensitive
clusters we identified, it is not our goal to quantify or exhaustively enumerate
this phenomenon. Instead, we anecdotally refer to a few representative examples
that highlight this dimension of privacy leakage. In one case, the user 
expressed negative feelings about his/her doctor, while the GPS coordinates 
place the user in the office of a mental health professional. In another example,
the user complained about some blood tests, while being geo-located at a rehab center.
Also, geotagged religion-based tweets 
can reveal the type of that place of worship (e.g., mosque, synagogue)
and may even point to a specific denomination. However, even if users are cautious and nothing sensitive 
is disclosed in the tweets, the location information obtainable with our duration-based approach 
can result in significant privacy loss.

\subsection{Impact of Historical Data}
\label{ssec:historical}

\begin{figure}[t]
	\centering
	\includegraphics[width=\columnwidth]{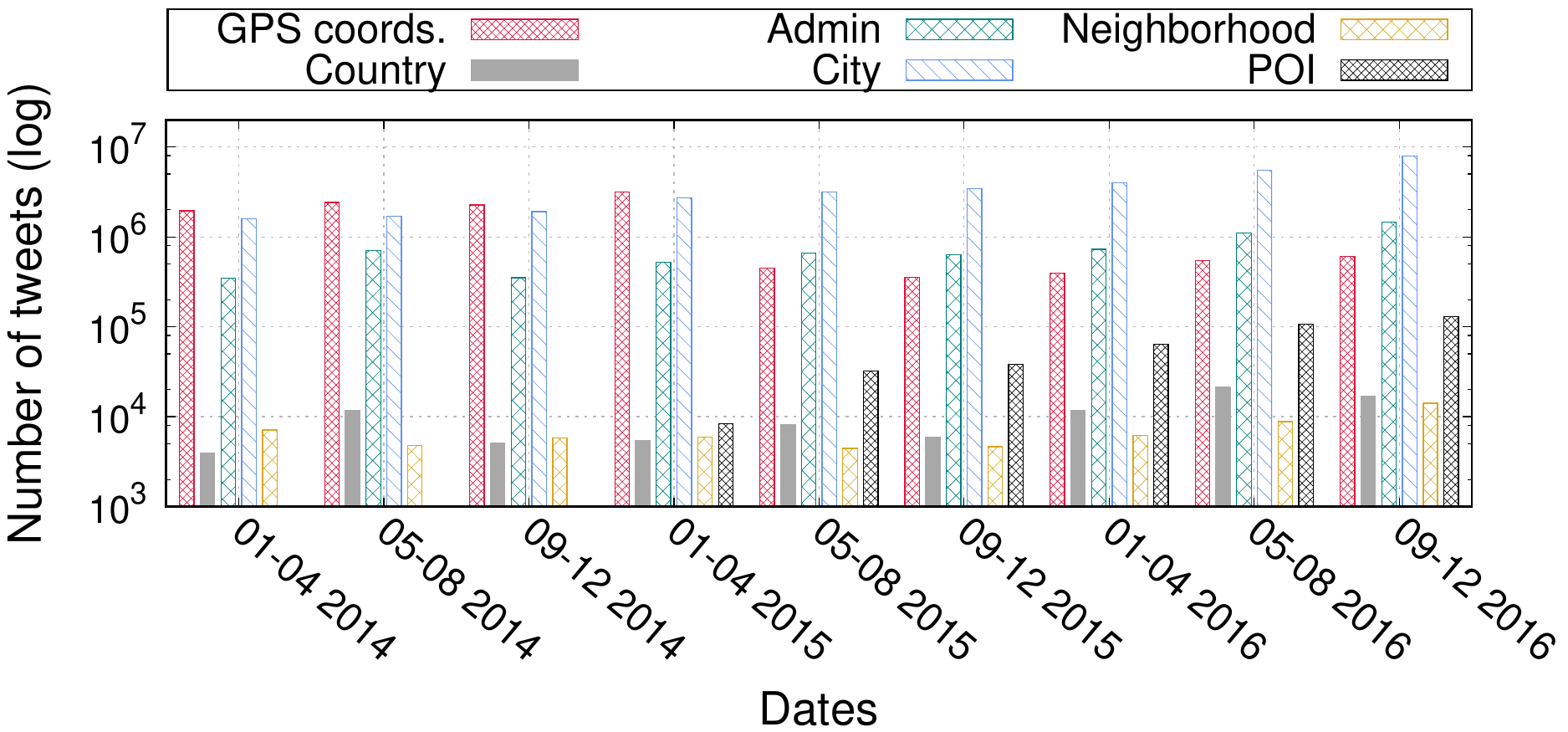}
	\caption{Granularity of location for all geotagged tweets between 2014-2016 in our entire dataset.}
	\label{fig:buckets}
\end{figure}

\begin{table}
	\caption{Tweets with GPS coordinates depending on Twitter's policy on
		including precise location metadata.}
	\label{tab:twitter_evolution}
	\centering
	\small
	\begin{center}
		\resizebox{\columnwidth}{!}{
			\begin{tabular}{lcc}
				\toprule
				\textbf{Dataset} & \textbf{Before 4/2015} & \textbf{After 4/2015}\\
				\midrule
				\rowcolor{Gray}
				All tweets & 24.98\% & 1.35\% \\
				Coarse-grained tweets & 99.9\% & 2.85\% \\
				\bottomrule
			\end{tabular}
		}
	\end{center}
\end{table}

During our analysis we found that Twitter app versions released prior to April 2015
automatically include GPS coordinates in tweets tagged with a coarse location.
The resulting tweets have both
a coarse-grained label (e.g., city) and GPS coordinates in their metadata.
Furthermore, that
information is not visible in the app or the web version. Thus, users
are completely oblivious to the public availability of this sensitive information. In newer versions
users have to explicitly opt to include GPS information on a per-tweet basis.
The apps with a more privacy-respecting behavior were released on
April 15\textsuperscript{th} for iOS and the 20\textsuperscript{th} for Android.
Nonetheless, the historical metadata collected from the prior versions \emph{remains
publicly accessible} through Twitter's API.

\textbf{Unavoidable privacy leakage.} 
As shown in Figure~\ref{fig:buckets} user behavior
changes after April 2015, with far fewer tweets with 
precise location, and users tagging tweets with the newly introduced point-of-interest (POI)
that denotes locations of varying granularity.
However, throughout the entire lifespan of our dataset,
users keep posting tweets tagged with a coarse-grained location, further indicating that
they are interested in enriching the context of their tweets, but not always willing
to disclose their exact whereabouts.
Table~\ref{tab:twitter_evolution} shows that there is significant
change, with a 35-fold reduction in the ratio of tweets that contain GPS
coordinates after the release dates of the apps with the privacy-respecting
approach.  Since we do not have the ability to detect each user's app version,
we first separate the data on the official release date for each platform
(i.e., we take into account if the user is on Android or iOS). While some users
may have delayed updating their app, that would only increase the ratio of
tweets with GPS; thus, the actual reduction of tweets with GPS is even higher,
further highlighting the unavoidable privacy violation that users faced due to
Twitter's poor handling of location data. 
While we expected that all
coarse-grained tweets from before 04/2015 would contain GPS coordinates, we found that $\sim$0.1\%
do not. These were all from before 08/2010 indicating the point in time when Twitter started
the practice of appending GPS data to coarse-grained tweets. Consequently this privacy-invasive policy persisted
for almost 5 years until Twitter gave users greater control over the location information they exposed.
Nonetheless, users with older devices or versions of the app are still exposing this type of data,
while all users' data remains accessible online.

\begin{figure}[t]
	\centering
	\includegraphics[width=1\columnwidth]{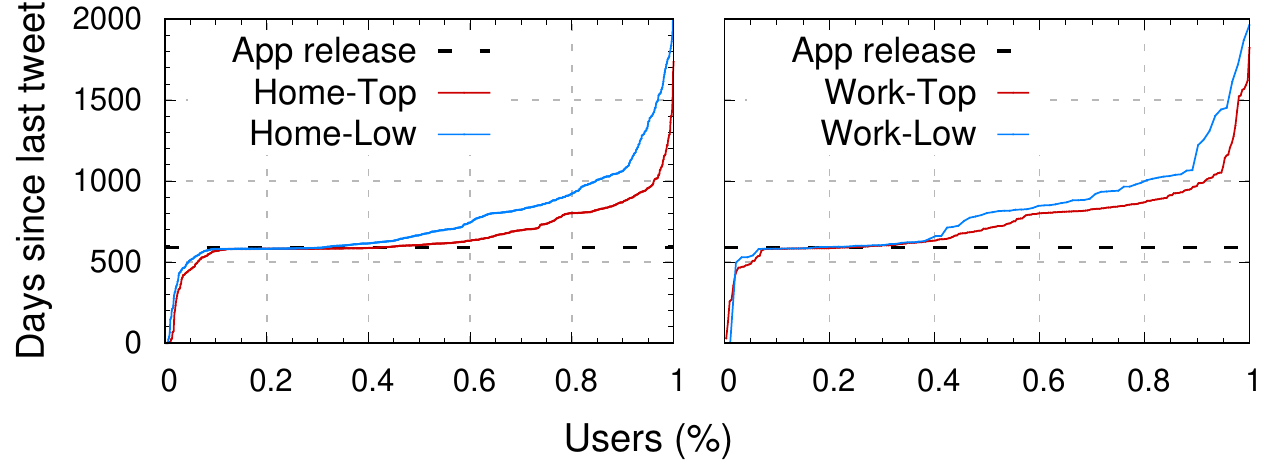}
	\caption{Days passed since the most recent tweet.} 
	\label{fig:freshness}
\end{figure}

\textbf{Historical data.} We explored the impact of Twitter maintaining and publicly 
sharing historical location metadata, by calculating how many users would remain vulnerable
if GPS coordinates were not included in coarse-grained tweets.
In Figure~\ref{fig:freshness} we first look at the number of days that have passed 
since the last tweet from a home/work location. 
We find that 56.57\% and 68.45\% of the users
posted their last tweet from home right before the release of the newer app version,
a large percentage around the dates of the app release, and only a small
number after that. As we do not have information regarding the 
date each user installed the newer app version on their device, we cannot
know the exact numbers. However, it is evident 
that the majority of users stopped posting tweets with precise 
location information from their home and work locations.

To further investigate how users' behavior changed since
Twitter changed its policy, we identified the
users that have posted tweets with coordinates after the app release 
date (and the following weeks) and ran \system only on the 
tweets posted after those dates. As can be seen in Table~\ref{tab:after_rel},
as users started updating their apps, the number of users posting tweets
with precise location drops rapidly. Indicatively, only 15.43\% and 11.12\% 
of the users in the two datasets continued posting such tweets
four weeks after the release of the new app.
When using only tweets posted at least four weeks after the app release
we were able to correctly identify the home of 7.34\% and 6.39\% of the 
users that are identified when all data is used.

Regarding the ``freshness'' of historical data, it is important to note that even 
if some of the users' key locations have changed (e.g., a user has since moved to a different home), users can still be 
identified by that data, and the inferred sensitive information does not ``expire''.
The sensitive user traits, actions or beliefs that can be inferred by the three categories
that we explore will still characterize the users regardless of the current location 
of their home or workplace. Even for ephemeral characteristics that no longer hold true,
exposure of that sensitive information can still affect 
users (e.g., certain cured medical issues remain social taboos).
As such, given the adage that ``the Web never forgets'', Twitter's
invasive privacy policy cannot be dismissed as a case of 
a vulnerability that has been fixed. As long as this historical data
persists online, users will continue to face the significant privacy 
risks that we have highlighted in this paper.

\begin{table}
	\caption{Home inference using geotagged tweets posted after the new geolocation policy of Twitter.}
	\label{tab:after_rel}
	\centering
	\small
	\begin{center}
		\begin{tabular}{lcccc}
			\toprule
			\textbf{Dataset} & \textbf{Date} & \textbf{Users} & \textbf{Homes} & \textbf{Coverage} \\
			\midrule
			\rowcolor{Gray}
			Home-Top & Release &602 & 333 & 35.96\% \\
			Home-Top & +4 Weeks & 155 & 68 &  7.34\% \\
			\midrule
			\rowcolor{Gray}
			Home-Low & Release & 394 & 239 & 24.66\% \\
			Home-Low & +4 Weeks & 116 & 62 &  6.39\% \\
			\bottomrule
		\end{tabular}
	\end{center}
\end{table}

\subsection{Performance evaluation}

\begin{figure}[t]
	\centering
	\includegraphics[width=0.9\columnwidth]{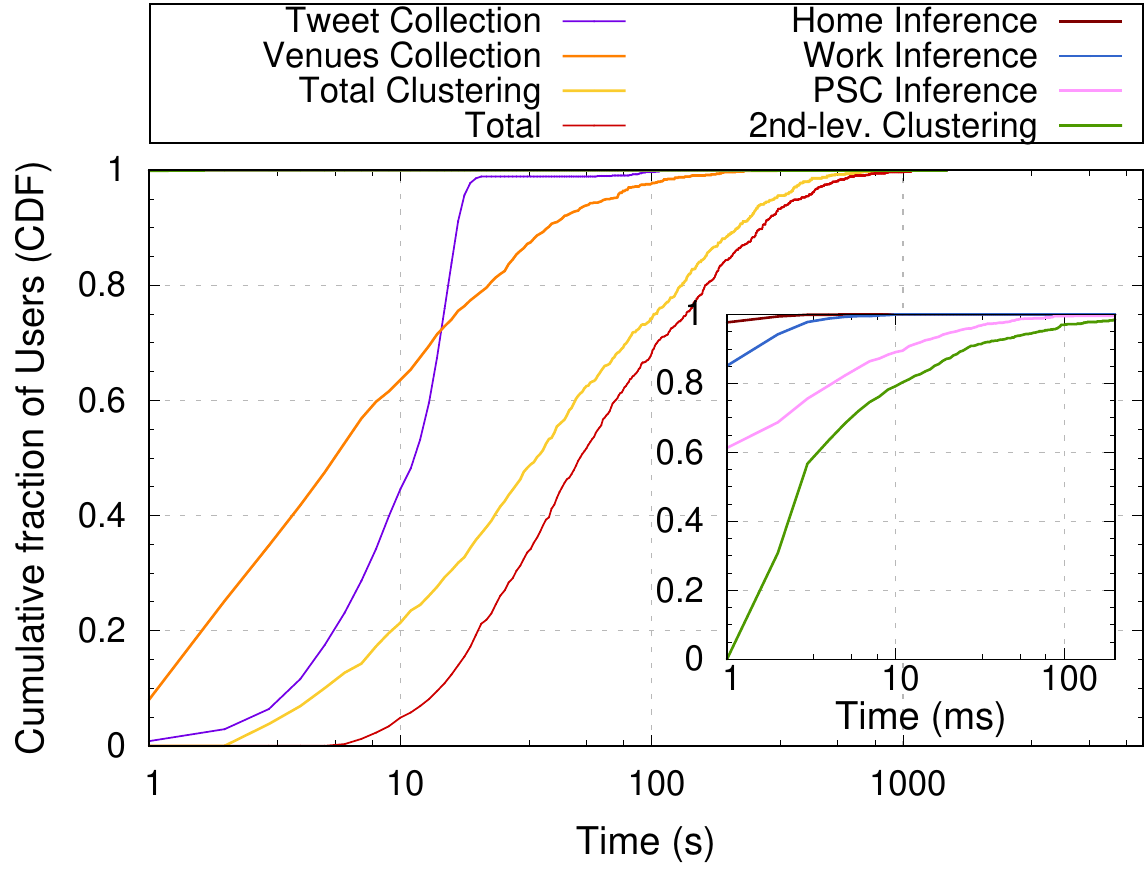}
	\caption{Time required by \system for each phase of the
		process. The reported numbers are for a subset of 1,000 users randomly
		selected from all the users with geotagged tweets.}
	\label{fig:performance_evaluation}
\end{figure}

To evaluate the performance of our system we selected 1000 
users randomly from all the users with geotagged tweets and measured the time 
required by each module of \system, as well as the total time, for 
completing the entire process. As expected, this time 
depends on the number of tweets and clusters each user has;
as such we randomly chose the users to reflect a representative distribution. 
As shown in Figure~\ref{fig:performance_evaluation}, the most 
time demanding operations are those of collecting a user's 
tweets, collecting PSCs, and the first-level clustering
all of which rely on the use of third-party APIs 
(i.e., communication over the network, rate limits, etc.).
On the other hand, the time required for the other steps
are in the order of milliseconds,
which can be considered as negligible. 

Using a commodity desktop,
\system requires less that 12 seconds for collecting all the tweets of roughly
half the users, and less than 20 seconds for around 98\% of the users.
Furthermore, for the collection of PSCs it takes up to six seconds for half of
the users, and more than 29 and 66 seconds for 15\% and 5\% of the users
respectively. For the process of clustering, our system takes up to 35 seconds
for about 50\% of the users, and more than 164 and 305 seconds for 15\% and 5\%
of the users. To that end, when considering the total time spent, \system
takes less than 52 seconds for half of the users, and more than 207 and 385 for
15\% and 5\% of them (users with a very large number of tweets and clusters).
Our system can complete the whole process in less than a minute for half the
users, while approximately 95\% of them can be processed within six minutes.
This highlights the severity and scale of the privacy threat we have explored,
as adversaries could trivially run such attacks for a massive number of users.

\section{Discussion and future work}
\label{sec:discussion}

\textbf{Twitter privacy leakage mitigation.} The pitfalls of location-sharing have long troubled researchers.
And while our work demonstrates the extent of the risks users face, it also
highlights an important aspect of the issue that, to our knowledge, has not been explored
before. While previous work has mainly focused on users knowingly or inadvertently sharing 
location data in social platforms,
we also identified an inconspicuous form of privacy leakage that is invisible to users.
Even though Twitter has since opted for a more privacy-oriented policy where
users have to explicitly choose to append GPS coordinates in tweets, the
availability of historical metadata severely undermines the benefits of this
more recent approach. Apart from the fact that after users are given the choice
they are 18.5 times less likely to include GPS coordinates, $\sim$93\% of the
users identified by \system are due to the historical tweets geotagged by
Twitter.  These findings underline the risks of web services \emph{publicly
over-sharing data} through their APIs, which poses an alarming flip side to the
common problematic behavior of over-collecting data~\cite{roberds2009data}.

We found that Twitter mentions this behavior~\cite{twitter_location},
and describes the process for removing location data~\cite{twitter_settings}.
However, they explicitly warn users that
``deleting location information on Twitter does not guarantee the information will be removed from 
all copies of the data on third-party applications or in external search results''. As data brokers 
continuously collect and sell Twitter data, even if users do remove all location metadata from their 
tweets, it is not necessary that those changes will be reflected in the versions maintained by others.
As users continue to tweet the privacy threats posed by historical data may 
naturally degrade through time, as less and less of those geotagged tweets will be obtainable through 
the free public API (which sets a limit for the 3200 most recent tweets). Interestingly this degradation
will occur faster for more prolific users.
However, it is important to note that third-party copies of that data will still exist,
while more resourceful or motivated attackers will still be able to utilize Twitter's paid APIs that 
give access to a \textit{complete archive} of users' tweets.

Ideally, our study will motivate services 
and act as a deterrent against publishing sensitive metadata not explicitly broadcast by users
(interestingly, a recent study explored how other types of metadata can
uniquely identify a user~\cite{Perez:2018}).
While the availability of public Twitter data has facilitated innovative and impactful research,
the privacy threats that users face remains an important issue. This is further exacerbated by the
significant ramifications for users that rely on the pseudonymous nature of
Twitter~\cite{184469}.
Overall, while Twitter offers a partial solution
for mitigating the privacy risks of this (historical) data, 
there exists no foolproof course of action for completely eradicating this threat.

\textbf{Applicability of \system.} 
The techniques used by our system for inferring users' home and work locations are not tied to 
Twitter, but can be readily applied to any (sparse) location dataset that contains periodic entries of 
GPS coordinates and timestamps. For datasets that contain very frequent snapshots of a user's 
location (e.g., continuously collected every couple of seconds), a simple form of sampling 
should be sufficient for reducing the computational overhead that can arise from a massive
number of data points. And while our content-based technique for inferring sensitive locations
is not applicable to every location-based service as it relies on the tweets' content,
our duration-based technique can also be applied to any location dataset.
However, it is important to stress that our evaluation focuses solely on Twitter
and the exact effectiveness of our techniques on other data sources remains unexplored.

\textbf{\system adoption.} Recent headlines regarding third parties harvesting personal user information
in services like Facebook~\cite{ny_fb} have reignited the public discourse over user privacy and data protection.
Facebook has announced plans for offering users more control over their data~\cite{reuters_facebook} and
Twitter is aiming for increased transparency due to the new GDPR requirements~\cite{twitter_gdpr}.
As such, it is evident that there is need for tools and techniques that can intuitively inform users
about what data of theirs is exposed. And while certain cases of data exposure
can be self-evident, sensitive information inference may be less obvious. To that end, \system can 
be incorporated by any location-based service or social network for clearly notifying users of such
exposure. For services that do not obfuscate locations,
users can also explore user-side location-obfuscation tools like LP-Doctor~\cite{kassem15anatomization}
or an app for location spoofing~\cite{polakis:acsac13}.

\textbf{POI.} While we find a massive reduction in tweets with GPS coordinates after the release of
the less privacy-invasive Twitter apps, there is a considerable number of tweets with a
``point of interest'' label. These labels cover areas from very coarse (e.g., ``Central Park'') to
fine-grained (e.g., ``Starbucks''). However, without the GPS coordinates it becomes much 
harder to place users at those locations, since the tweet may be \emph{about
that place} instead of \emph{at that place}~\cite{twitter_api_object}.
We consider the exploration of POI-based privacy leakage as future work.

\textbf{Ethical considerations and disclosure.}
As is the case with any study that explores user privacy, it is important to address the ethical implications
of our work. A precise description of our study and experimental protocol were submitted to and approved for 
exemption by our university's Institutional Review Board (IRB). Moreover, apart from only collecting 
\emph{publicly available data} offered by the official Twitter API, all
usernames were removed during the manual annotation process. This ensured that the
authors would not be able to identify/deanonymize any users. At the
same time, all collected data and results from our analysis were stored on
machines with up-to-date software, encrypted hard drives, where access was strictly limited to the
authors and only possible from two white-listed internal IP addresses using
authorized SSH keys. We believe that our research presents minimal risk while having the potential 
for significant benefits to users; we have submitted a report to Twitter outlining our research and findings,
and substantiating the need to purge this historical data. Moreover,
we have deleted all the results of our analysis, as well as the ground truth dataset and the entirety of the 
collected Twitter data, to eliminate any potential of future privacy risks to the users.

By reporting the accuracy of our location identification techniques we can alert users of
the true extent of the privacy threats they face. Also, by educating users on the risks of location sharing we 
hope to motivate even more privacy-conscious behavior.
Finally, demonstrating the practicality and feasibility of impactful attacks can
incentivize services to incorporate privacy-preserving techniques
for collecting and sharing location data
(e.g.,~\cite{Gruteser:2003,Andres:2013,kassem15anatomization})
and offer the functionality of \system as an intuitive auditing tool for users to assess 
their level of privacy exposure.

\section{Related Work}
\label{sec:related}

Prior work has proposed approaches for identifying home and work locations
that range from inspecting social graphs, to studying 
check-ins and precise geolocation data (a survey can be found in~\cite{zheng:2018}). As certain users do not
geotag their tweets, previous work has also tried to infer home
locations based on tweet content~\cite{cheng_2010,mahmud_2014,ryoo_2014}
or other information like social ties~\cite{gu_2016,chen:2016} or check-in behavior~\cite{li_2012}. However, 
these studies can only infer key locations with a very coarse granularity. 
Furthermore, the inference of sensitive information from other location data points
has not been explored.

\textbf{Location inference.} In a study that investigated mobility patterns, Cho et al.~\cite{cho_2011} 
considered geographic cells of 625km\textsuperscript{2}, and considered a 
home location to be the average position of
the cell with the most check-ins. Pontes et 
al.~\cite{pontes} used Foursquare check-ins and correctly 
inferred the home city of $\sim$78\% of users. By considering that
users are located at their home at night and near their office during 
working hours, Liu et al.~\cite{liu_2014} 
identified the key locations of 68\% of the users within
a distance of 2.5 km.
Efstathiades et 
al.~\cite{efstathiades_2015} followed an approach similar to Liu et al.'s for detecting users' 
home and work at a postcode granularity 

using three Twitter datasets from the Netherlands, London, and 
the Los Angeles county. Given that the average size of a postal code area
in LA is approximately 14.66km\textsuperscript{2} and includes over 37K residents (our calculations 
are based on data from~\cite{city_data}), it is evident that postcode-level granularity 
is still very coarse.

Furthermore, the work by Isaacman et al.~\cite{Isaacman_2011} leveraged
cellular network data to identify important user locations, based on different
factors such as the number of days a cluster of cell towers was contacted, and
the duration and number of such events, during \textit{home hours} (7pm-7am)
and \textit{work hours} (1pm-5pm on weekdays) i.e., the number of times a cell
tower was contacted in the corresponding set of hours. Then, by employing
logistic regression and setting several thresholds derived from 18 training
volunteers, the authors estimated the likelihood of a cluster being
\textit{important}. Out of these clusters, again relying on logistic
regression, they would pick as home the cluster deemed most significant by the
\textit{home hour events} factor (i.e. the one with the largest number of
events during the home hours), and as work the cluster with a high \textit{work
hour} and a low \textit{home hour} score. In a similar
setting,~\cite{Hightower_2005} and~\cite{Kang_2004} relied on GSM and WiFi
transmission beacons to identify significant user locations, but these two
works did not attempt to label the identified locations as home or work.
However, in reality, approaches that rely on that type of data (i.e., cellular
data) are limited in terms of granularity that can be achieved. Indicatively, the authors
of~\cite{Isaacman_2011} estimated that the median distance error for their home
and work inference approach is around 1.44Km and 1.33Km. This is expected
though, considering that the distance between the actual home and work
locations (reported by the study's participants) and their nearest cell towers
are 0.98Km and 0.8Km for home and work, respectively.

While our approach for detecting home and work locations 
is based on spatiotemporal analysis, similarly to the previous
work in the area, 
we are the first to propose an approach that considers the vertical ``widespreadness'' of users' 
activity for detecting the home location, as well as a dynamic and adaptive approach for detecting 
the user's work location after estimating working hours (night shifts etc.).

Apart from \system outperforming previously proposed techniques (as shown in
Section~\ref{sec:evaluation}), our experimental evaluation and comparative study
was conducted on a ground truth dataset that is significantly more complete and
fine-grained than the datasets used in prior studies. In detail, the ground
truth constructed in~\cite{efstathiades_2015} was at a postcode level,
while Cheng et. al~\cite{cheng2011exploring} did not actually verify their home selection
with some form of ground truth.  Similarly, in~\cite{cho_2011} the authors
constructed a ground truth using 25x25km cells and stated that ``manual
inspection shows that this infers home locations with 85\% accuracy'' but did
not include more details on how that was done.  Furthermore, the datasets
in~\cite{krumm2007,hu_2015,lin_2017} all contained only home locations, with the
dataset  by Lin et al.~\cite{lin_2017} being based solely on the visual inspection of the
GPS data points. While in~\cite{hu_2015} the authors also relied on manual
inspection of tweet content for identifying the home locations, that process 
was conducted by Amazon Mechanical Turk workers who were only shown a subset
of five tweets from a cluster, whereas our manual inspection was conducted collectively 
on the entirety of tweets assigned to each cluster.

\textbf{Location and de-anonymization.} The problem of identifying key locations has also been explored 
in different settings, e.g., using continuous GPS data collected from
receivers in cars~\cite{krumm2007} or wearables~\cite{liao2005location}. Golle and Partridge 
built upon these findings and explored how users can be 
identified from different granularities of anonymized census data~\cite{golle2009anonymity}.
De Montjoye et al.~\cite{de2013unique} 
explored the uniqueness of user mobility patterns in a 15-month dataset
for 1.5M people, and found that four coarse-grained spatiotemporal points can uniquely differentiate 95\% of 
the individuals within the set. Previously, Chong et al.~\cite{song2010limits} reported a 93\% predictability 
in mobility by measuring the entropy of users' trajectories. Rossi et al.~\cite{RosMus15_spatiotemporal}
demonstrated the feasibility of identifying users within mobility traces by using movement data including speed, direction
and distance of travel and found that as little as two location points may be sufficient to uniquely identify users.
Zang et al.~\cite{Zang_2011} leveraged an anonymized three-month dataset of mobile call records, consisting of
roughly 25M users, and discovered that as few as the \textit{top-2} most frequented locations can re-identify a
user, even in varying granularities, such as call sector, cell, zip code level etc.

\textbf{User behavior.} Prior work also explored how users interact with or disclose location data,
and the feasibility of social-tie inference.
Liccardi et al.~\cite{liccardi2016} explored how different ways of visualizing 
data affected users in inferring the type of a location (home, work, etc).
Ahern et al.~\cite{ahern2007over} investigated
how users select the privacy settings for uploaded photos,
and found that users are more likely to set as private photos that are 
taken at frequently photographed locations while 
tending to set photos from less frequented locations public.
Consolvo et al.~\cite{consolvo2005location} found that users were willing 
to disclose exact locations, but that study focused on a different 
setting where users were asked about sharing information with friends, 
family and colleagues. 
Tang et al.~\cite{tang2010rethinking} identified how
users adapt their location sharing behavior 
and explored the different ways and levels of granularity at which 
users decide to share their location under different hypothetical scenarios.
Cheng et al.~\cite{cheng2011exploring} conducted a large-scale study of location data and studied mobility patterns 
and the correlation between check-ins and message content and sentiment. 
Sadilek et al.~\cite{sadilek_2012} proposed a probabilistic human mobility 
model for predicting users' social links and locations. That model considers 
users who disclose GPS coordinates as noisy sensors for inferring the location 
of their friends. 
Several other works (e.g.,~\cite{Crandall_2010, Pham_2013, Scellato_2011, Zhang_2015}) leverage spatiotemporal data
for inferring social ties among users.
More recently, Backes et al.~\cite{backes_2017} developed
an attack for inferring social relationships from mobility data, 
which constructs and compares user mobility profiles in an 
unsupervised setting. Working on the opposite perspective, 
Aronov et al.~\cite{aronov} leverage relationships and
additional information such as co-location of events
to infer other potentially visited locations.

\textbf{Aggregate Location Data.} 
	Prior work has also studied how aggregate location time-series can lead to significant privacy loss.
	Pyrgelis et al.~\cite{Pyrgelis_2018} presented a novel methodology to study
	membership inference on such data, by formalizing the problem as a distinguishability game,
	and showed that such attacks are indeed feasible and can lead to significant
	privacy loss depending on different factors, such as the adversary's prior knowledge, aggregation group sizes etc. Shokri et al.~\cite{shokri} investigated membership machine learning based membership inference, and used a dataset of Foursquare check-ins to evaluate their attack.
	In another work, Pyrgelis et al.~\cite{Pyrgelis_2017} studied how 
	aggregate location data can be leveraged to localize
	and even profile individual users, under different adversarial knowledge scenarios. Finally, Xu et al.~\cite{Xu_2017},
	demonstrated that it is possible to recover individual user trajectories from aggregate location data with high accuracy,
	based on the key assumptions that users' daily trajectories tend to be regular but also differ
	significantly among users. They also highlight that due to the high uniqueness of the recovered trajectories further
	attacks, such \textit{re-identification}, are enabled.

\section{Conclusions}
\label{sec:conclusions}

We have investigated the privacy threats that arise from precise 
location (meta)data being publicly accessible in Twitter's API. By developing novel techniques 
for identifying a user's exact home and work location, and 
inferring sensitive information through the reconstruction of a user's location history,
\system highlights the true extent of the risk of exposing such information.
To make matters worse, our experimental evaluation revealed how Twitter's invasive
policy of including precise location data in previous app versions has 
significant implications, as it results in an almost 15-fold increase in the number 
of users whose key locations are successfully identified by our system.
Given that users are most 
likely oblivious to this privacy leakage,
it is important to shed light on this privacy-invasive
practice. We hope that our work will serve as a cautionary tale, 
equipping users with the means to manage their personal 
data and avoid the risks of public exposure.

\section*{Acknowledgments}

We would like to thank the anonymous reviewers for their valuable feedback. Special thanks to 
our shepherd Emiliano De Cristofaro, for all his help. 
The research leading to these results has received funding from European Union's
Marie Sklodowska-Curie grant agreement No 690972 and Horizon 2020 Research \& 
Innovation Programme under grant agreement No 780787 and No 740787,
and the Defense Advanced Research Projects Agency (DARPA) ASED Program
and the Air Force Research Laboratory (AFRL), under contract
FA8650-18-C-7880.
This paper reflects only the view of the authors and the funding bodies are not 
responsible for any use that may be made of the information it contains.

\bibliographystyle{IEEEtranS}
\bibliography {paper.bib}

\end{document}